\documentclass[a4paper,11pt]{article}
\usepackage{jheppub} 
\usepackage{lineno}
\pdfoutput=1 
\usepackage{subcaption}
\usepackage{graphicx,color}
\usepackage{enumitem}
\usepackage{braket}
\usepackage{slashed}
\usepackage[utf8]{inputenc}
\usepackage{hyperref}
\usepackage{float}
\usepackage{caption}
\usepackage{makecell}
\usepackage{epsfig}
\usepackage{amsmath,amssymb}
\usepackage{hyperref}
\usepackage{diagbox}
\usepackage{lscape}
\usepackage{bbold}
\usepackage{algorithm}
\usepackage{algpseudocode}
\usepackage{amsmath}
\usepackage{amsthm}
\usepackage{comment}
\usepackage{booktabs}
\usepackage{siunitx}
\usepackage{mathrsfs}
\usepackage{soul}
\usepackage{listings}
\usepackage{array}
\usepackage{multirow}
\usepackage{booktabs} 

\hypersetup{
	colorlinks=true,
	citecolor=black,
	filecolor=black,
	linkcolor=blue,
	urlcolor=blue
}

\usepackage{dsfont}

\newcommand{\comments}[1]{}

\def\bel#1{\begin{equation} \label{#1}}

\def\ED3{{\scriptscriptstyle ED3}}

\newcommand{\beq}{\begin{equation}}  \newcommand{\eeq}{\end{equation}}
\newcommand{\bal}{\begin{aligned}}   \newcommand{\eal}{\end{aligned}}
\def\ov{\overline}
\newcommand{\bmat}{\left(\begin{array}}
\newcommand{\emat}{\end{array}\right)}

\newcommand{\cN}{\mathcal{N}}

\newcommand{\cI}{\mathcal{I}}

\newcommand{\cR}{\mathcal{R}}

\newcommand{\cM}{\mathcal M}

\newcommand{\ee}{\text{e}}

\usepackage{comment}
\usepackage[normalem]{ulem}

\def\bZ{\mathbb{Z}}
\newcommand{\I}{\text{i}}

\newcommand{\kom}{\, ,\quad }

\newcommand*{\p}{\mathop{}\!\mathrm \partial}

\newcommand{\bC}{\mathbb{C}}

\usepackage{orcidlink}

\title{Flux Vacua Near the Boundary of Large Complex Structure}

\author[a,b]{Aman Chauhan$\,$\orcidlink{0009-0009-0596-0928},}
\author[c,d]{Michele Cicoli$\,$\orcidlink{0000-0003-1709-5651},}
\author[a,b,e]{Anshuman Maharana$\,$\orcidlink{0000-0001-8627-6398},} 
\author[f]{Pellegrino~Piantadosi$\,$\orcidlink{0009-0000-5284-9485}}

\affiliation[a]{\footnotesize Harish-Chandra Research Institute, Chhatnag Road, Jhunsi,
Prayagraj, Uttar Pradesh 211019, India}

\affiliation[b]{Homi Bhabha National Institute, Training School Complex, Anushakti Nagar, Mumbai
400094, India}

\affiliation[c]{\footnotesize Dipartimento di Fisica e Astronomia, Università di Bologna, via Irnerio 46, 40126 Bologna, Italy}
\affiliation[d]{\footnotesize INFN, Sezione di Bologna, viale Berti Pichat 6/2, 40127 Bologna, Italy}
\affiliation[e]{\footnotesize Leinweber Institute for Theoretical Physics, Randall Laboratory of Physics,
University of Michigan, Ann Arbor
450 Church St, Ann Arbor, MI 48109-1040, USA}
\affiliation[f]{New York University Abu Dhabi, PO Box 129188, Saadiyat Island, Abu Dhabi, UAE}

\emailAdd{amanchauhan@hri.res.in}
\emailAdd{michele.cicoli@unibo.it}
\emailAdd{anshumanmaharana@hri.res.in}
\emailAdd{pellegrino.piantadosi@nyu.edu}

\abstract{Three-form fluxes generate a potential for the complex structure moduli in type IIB string compactifications on Calabi-Yau threefolds. In the large complex structure patch, the potential consists of a perturbative contribution and a convergent series of instanton corrections. We present examples of vacua near the boundary of large complex structure, where the minimum of the potential is essentially determined by the perturbative contribution and the first instanton correction, while the effect of all higher instantons is negligible. This phenomenon occurs when fluxes are such that the magnitude of the perturbative potential in certain directions in moduli space is of the same size as the first instanton contribution. Analysing an ensemble of flux vacua, we find that this phenomenon is statistically quite common. We also discover more subtle phenomena where instanton terms affect the multiplicity of the solutions and induce monodromy shifts.}

\preprint{LITP-26-14}

\begin{document}
\maketitle
\flushbottom

\section{Introduction}
\label{sec:intro}

Our current understanding of string theory primarily relies on expansions about asymptotics in moduli spaces. As a consequence, the best understood vacua are those located in these asymptotic limits, where it suffices to consider the leading terms in the effective action. On the other hand, the construction of viable string theory vacua (which can describe our Universe) requires moving away from the asymptotics of moduli spaces. This leads to challenges associated with control over the expansions as set by the Dine-Seiberg argument \cite{Dine:1985he}\footnote{For a detailed discussion and avenues for addressing the issue see e.g. \cite{Cicoli:2023opf}.}. Therefore, any study of moduli stabilisation in string theory requires a detailed analysis of the expansions that control the effective action and, consequently, the resulting potential. There is steady progress in this direction \cite{McAllister:2023vgy}. Gaining a complete understanding in a setting where all moduli are stabilised remains an open question as we still lack computational control over all the terms in the effective action (perturbative and non-perturbative).

This article will study this question, specialising 
to complex structure moduli stabilisation in IIB flux compactifications, working in the large complex structure patch. This specialisation gives us two important advantages. Firstly, the expansion is known to all orders and is convergent (in the large complex structure patch). Secondly, ref. \cite{KlemmKreuzerInstanton} together with recent computational advances \cite{Demirtas:2022hqf, Demirtas:2023als}, allow for an explicit computation of the terms to very high order. We will confine ourselves to a simple two-modulus Calabi-Yau setting so that a detailed analysis can be performed.

Recall that in the large complex structure limit, the flux superpotential consists of perturbative terms which are polynomial in the complex structure moduli and instanton terms which have exponential dependence on the moduli. In this setting, the simplest vacua are those in the deep interior of the large complex structure region where the minimum is well described by the perturbative contributions with the instanton terms leading to small corrections (these have been well studied in the literature, see \cite{Martinez-Pedrera:2012teo, Dubey:2023dvu, Chauhan:2025rdj}).

In this paper, we study instead flux vacua near the boundary of large complex structure where the instanton terms
have a non-trivial effect on the minimum while the expansion is still under control. In the simplest class of such vacua, the minimum is well described by the perturbative contribution and the first instanton correction, with all other terms in the instanton expansion negligible. As one would expect in such a situation, key features of these vacua are directions in moduli space where the perturbative prepotential and the first instanton correction make comparable contributions to the potential (which are larger than those arising from other instanton terms). The interplay between two effects makes this possible. Firstly, the fluxes are such that the perturbative contribution is lower than the naive expectation. Secondly, the values of the complex structure moduli at the minimum are only moderately large. Consequently, the contribution from the first instanton correction is not negligible, but comparable to the lowered perturbative contribution. At the same time, the values of the complex structure moduli are large enough to make the contribution of all higher instantons negligible. Thus, the first instanton correction leads to a large displacement of the perturbative minimum. A statistical analysis of flux ensembles reveals that this class is not uncommon. We also find more intricate phenomena where the inclusion of instanton terms affects the multiplicity of the flux vacua and induces
monodromy shifts. 

The structure of the paper is as follows. In Sec.~\ref{sec:typeIIB} we review the framework of type IIB flux compactifications at large complex structure relevant for our analysis and establish the notation and the conventions used throughout. In Sec.~\ref{sec:explicit_analysis} we present a detailed application to a two-modulus Calabi-Yau orientifold, providing explicit examples of flux vacua near the boundary of the large complex structure region of moduli space where the first non-perturbative effects are important. We investigate the statistical properties of this ensemble of vacua and generic non-perturbative features in Sec. \ref{sec:pvsnp}. We summarise and conclude in Sec.~\ref{sec:summary}. App. \ref{app:gv_invariants} shows the non-perturbative prepotential up to degree $10$, while the numerical study of the convergence of the instanton series is discussed in App.~\ref{App:convergence_region}.

\section{Type IIB flux compactifications}
\label{sec:typeIIB}

In this section, we briefly discuss type IIB flux compactifications and collect results relevant for our purposes. This will also set conventions and notations that will be followed. Detailed reviews on the subject are \cite{Grana:2005jc, Douglas:2006es}. 

\subsection{Calabi-Yau compactifications at large complex structure}
\label{sec:lcs}

Let $(X_3,\widetilde{X}_3)$ be  mirror dual Calabi-Yau threefolds and $\sigma: X_3\rightarrow X_3$ a holomorphic and isometric involution of $X_3$ which transforms the holomorphic $3$-form  as $\Omega\mapsto -\Omega$. We will denote the corresponding O3/O7 orientifold compactification of type IIB superstring theory by $X_3/\sigma$. The associated effective supergravity theory preserves $\mathcal{N}=1$ supersymmetry in four dimensions.

Under the orientifold action, the cohomology groups $H^{p,q}(X_3)$ of the Calabi-Yau split into odd and even eigenspaces, $H^{p,q}_{\pm}(X_3)$.  Complex structure moduli surviving the projection are part of $\mathcal{N}=1$ chiral multiplets with multiplicity $h^{1,2}_-(X_3,\sigma)= \text{dim}\, (H^{1,2}_-(X_3))$. We shall denote them $z^i$, $i=1,\ldots,h^{1,2}_-(X_3)$. In this work, we remain agnostic about other moduli sectors. For simplicity, we will focus on cases where $h^{1,2}_+(X_3)=0$ i.e $h^{1,2}_-(X_3)=h^{1,2}(X_3)$.\footnote{Orientifolds with these properties can e.g.~be obtained systematically using techniques described in \cite{Moritz:2023jdb}, see also \cite{Jefferson:2022ssj}.}

One can introduce a symplectic basis of $\{\Sigma_{I},\Sigma^I\} \subset H_3(X_3,\mathbb{Z})$ together with the corresponding Poincaré dual forms $\{\alpha^I,\beta_I\}$. The \emph{periods} are defined  by integrating the holomorphic $3$-form $\Omega$ over these cycles, and collectively denoted  by a  vector $\Pi$:
\begin{equation}
\label{eq:PeriodVecGen}
X^I=\int_{\Sigma_{I}}\Omega=\int_{X_3} \Omega\wedge \alpha^I\, ,\quad \mathcal{F}_I=\int_{\Sigma^I}\Omega=\int_{X_3} \Omega \wedge \beta_I \, , \quad \Pi =\left (\begin{array}{c} \mathcal{F}_I \\ X^I \end{array}\right ) \, .
\end{equation}
The periods $X^I$ act as homogeneous complex coordinates on a local patch of the complex structure moduli space of $X_3$. Away from  $X^0=0$, one can introduce projective coordinates $z^i =X^i/X^0$, $i=1,\ldots,h^{1,2}(X_3)$, with  $\Omega$
normalised such that $X^0=1$. The dual periods $\mathcal{F}_I=\mathcal{F}_I(z)$ are  determined by a prepotential $F(z)$ through
\begin{equation}
    \mathcal{F}_i(z)=\partial_{z^i} F(z) \, ,\quad \mathcal{F}_0 =2F-z^i\p_{z^i}F\, .
\end{equation}

Our focus will be on the \emph{Large Complex Structure} (LCS) region of the complex structure moduli space $\cM_{\text{cs}}(X_3)$. Mirror symmetry maps this region of the moduli space (of type IIB reduced on $X_3$) to the \emph{large volume region} of type IIA compactified on the mirror dual Calabi-Yau $\widetilde{X}_{3}$. Using mirror symmetry, it can be shown that the prepotential $F(z)$ in the LCS region takes the form \cite{Candelas:1990rm,Ceresole:1992su, Candelas:1993dm,Hosono:1993qy,Hosono:1994av}
\begin{equation}
    \label{eq:prepotential}
    F(z)=-\frac{1}{6}\widetilde{\kappa}_{ijk}\, z^i\,z^j\,z^k+\frac{1}{2}a_{ij}\,z^i\,z^j+b_i\,z^i +\tilde{\xi} + F_{\text{inst}}(z).
\end{equation}
The parameters appearing above are given in terms of a basis of $(1,1)$-forms $J_{i}\in H^{1,1}(\widetilde{X}_{3},\bZ)$ and the second Chern class of the mirror manifold $\widetilde{X}_{3}$ denoted by $c_2(\widetilde{X}_{3})$. They are
\begin{align}
\widetilde{\kappa}_{ijk} &= \int_{\widetilde{X}_{3}} \, J_i \wedge J_j \wedge J_k\kom a_{ij} = \frac{1}{2}\int_{\widetilde{X}_{3}} \, J_i \wedge J_j \wedge J_j\,\text{mod}\,\mathbb{Z}\; ,\quad \nonumber\\
b_j &= \frac{1}{4!}\int_{\widetilde{X}_{3}} \,c_2(\widetilde{X}_{3}) \wedge J_j\kom \tilde{\xi} = \frac{\I}{2}\frac{\zeta(3)\, \chi(\widetilde{X}_{3})}{(2\pi)^3}\,.
\end{align}
The expansion \eqref{eq:prepotential} has
a finite radius of convergence~\cite{Hosono:1994av}  (see also \cite{Candelas:1994hw, Klemm:1999gm}); this serves as the definition of the LCS patch.

The non-perturbative contributions $F_{\text{inst}}$ in \eqref{eq:prepotential} arise from worldsheet instanton effects on the mirror dual and are given by~\cite{Hosono:1994av, Hosono:1994ax}
\begin{equation}
\label{eq:InstCorrections} 
    F_{\text{inst}}(z)=-\dfrac{1}{(2\pi \I)^{3}}\sum_{q \in \cM(\widetilde{X}_{3})}\, \mathscr{N}_{\tilde{\mathbf{q}}}\, \text{Li}_{3}\left (\ee^{2\pi \I\, q_i\, z^i}\right )\kom \text{Li}_{3}(x)=\sum_{m=1}^{\infty}\, \dfrac{x^{m}}{m^{3}}\, .
\end{equation}
Here, the sum is over the effective curves $q$ in the \emph{Mori cone} $\cM(\widetilde{X}_{3})$ of the mirror $\widetilde{X}_{3}$ and $\mathscr{N}_{\tilde{\mathbf{q}}}$ are genus-zero \emph{Gopakumar-Vafa (GV) invariants}
~\cite{Gopakumar:1998ii,Gopakumar:1998jq}.  
A systematic method for evaluating these invariants was developed in~\cite{Hosono:1993qy, Hosono:1994ax}. In practice, they can be calculated using the software package \texttt{CYTools}~\cite{Demirtas:2022hqf, Demirtas:2023als}. 

This provides a description of the moduli space of K\"ahler structures on $\widetilde{X}_{3}$, parametrised by a K\"ahler form $J$. K\"ahler cone computations can also be performed using \texttt{CYTools}. 

\subsection{Flux superpotential and supersymmetric vacua}
\label{sec:flux_vacua}

One can turn on background fluxes: $H_3$ and $F_3$ threading $3$-cycles of the Calabi-Yau. These are required to obey Dirac quantisation conditions. In terms of the above symplectic basis, we shall denote the quanta as follows:
\begin{equation}
(f_2)^I 
=\int_{\Sigma_{I}}F_3\, ,\quad (f_1)_I
=\int_{\Sigma^I}F_3\, ,\quad (h_2)^I
=\int_{\Sigma_{I}}H_3\, ,\quad (h_1)_I
=\int_{\Sigma^I}H_3\, ,
\end{equation}
and combine them into two integral flux vectors $f,h\in \mathbb{Z}^{2(h^{2,1}+1)}$
\begin{equation}\label{eq:fluxdef}
    f=\left (\begin{array}{c}
                f_{1} \\ 
                f_{2}
                \end{array} 
    \right ) \, ,\quad 
    h=\left (\begin{array}{c}
                h_{1}\\ 
                h_{2} 
            \end{array} 
    \right )  \kom f_{1},f_{2},h_{1},h_{2}\in\mathbb{Z}^{h^{2,1}+1}\, .
\end{equation}
The fluxes are constrained by Gauss's law tadpole condition for D3 charge, which reads:
\begin{equation}
     2\left(N_{\text{D3}}-N_{\overline{\text{D3}}}\right) + N_{\text{flux}} - Q_{D3} = 0\,,\label{eq:D3tadpole}
\end{equation} 
where $N_{\text{D3}}$ ($N_{\overline{\text{D3}}}$) is the number of spacetime-filling (anti-)D3-branes. Furthermore, we have introduced
\begin{equation}\label{eq:fluxtadpole}
     Q_{D3} =  \dfrac{\chi_f}{2}\, ,\quad N_{\text{flux}} =  \int_X H_3 \wedge F_3 = f^{\, T}\cdot \Sigma\cdot h\,,
\end{equation} 
in terms of the Euler character $\chi_f$ of the fixed locus of $\sigma$ in $X_3$. The tadpole cancellation condition \eqref{eq:D3tadpole} has to be satisfied in any consistent solution of string theory. Thus, if e.g. the D3-charge contribution from fluxes is such that $N_{\text{flux}}<Q_{D3}$, one needs to introduce spacetime-filling D3-branes. 

In the  effective four-dimensional $\cN=1$ supergravity theory, the tree-level K\"ahler potential $K$ for the complex structure moduli and the axio-dilaton is
\begin{equation}
\label{eq:TreeLevKP}
    K=-\ln( -\I\, \Pi^{\dagger}\cdot\Sigma\cdot \Pi ) - \ln\left(-\I({\tau}-\ov{\tau})\right) \kom \Sigma=\left (\begin{array}{cc}
        0 & \mathds{1} \\ [-0.2em]
        -\mathds{1} & 0
        \end{array} \right )\, .
\end{equation}
The F-term scalar potential is $V_F= V_{\rm flux}/\mathcal{V}^2$ where $\mathcal{V}$ is the  Calabi-Yau (dimensionless) volume in string units,  the flux potential is:
\begin{equation}
\label{eq:scalarpotential}
    V_{\text{flux}}= \ee^{K} \left ( K^{\tau\bar{\tau}}D_{\tau} W\, D_{\bar{\tau}}\overline{ W}+ K^{i\bar\jmath}D_{i} W\, D_{\bar\jmath}\overline{ W}\right )\, ,\quad D_I W = \partial_I W + (\partial_{I}K) W \,
\end{equation}
where $W$ is the Gukov-Vafa-Witten (GVW) superpotential~\cite{Gukov:1999ya}:
\begin{equation}
\label{eq:SupPotPerVec} 
    W=\int_{X_3} G_{3} \wedge \Omega = \left (f-\tau h\right )^{T}\cdot \Sigma\cdot   \Pi(z) \, .
\end{equation}
$W$ is protected from perturbative corrections by non-renormalisation theorems \cite{Giddings:2001yu, Burgess:2005jx}; it receives non-perturbative contributions from D-brane instantons, which we will ignore in what follows.  We also ignore perturbative corrections to the K\"ahler potential since they are expected to be subdominant in the large volume regime and when the string coupling is small.

Type IIB  string theory enjoys an $\text{SL}(2,\mathbb{Z})$ symmetry under which the axio-dilaton and $3$-form fluxes transform as
\begin{equation}
\tau \to  {{a \tau + b} \over {c\tau+ d}} \, ,\qquad \begin{pmatrix}
h \cr f
\end{pmatrix}
\to
\begin{pmatrix}
d & c \cr
b & a
\end{pmatrix}
\begin{pmatrix}
 h \cr f
\end{pmatrix} \, ,\qquad \begin{pmatrix}
a & b \cr
c & d
\end{pmatrix} \in SL(2,\mathbb{Z})\,.   
\label{eq:SLtrafo}   
\end{equation}
Under this transformation, the tadpole cancellation condition \eqref{eq:fluxtadpole} is unchanged, but the GVW superpotential \eqref{eq:SupPotPerVec} transforms non-trivially.
By performing $\text{SL}(2,\mathbb{Z})$ transformations successively, the axio-dilaton $\tau=c_0 + \I s$ can be made to take value in a \emph{fundamental domain} $\cM_\tau$ of $\text{SL}(2,\mathbb{Z})$.  We will choose this to be
\begin{equation}
\label{eq:FDAD}
   \cM_\tau = \biggl \{\tau=c_0+\I s\in\bC:\, |c_{0}|\leq 0.5\,\text{ and }\, |\tau|\geq 1\biggl \}\,.
\end{equation}
Additionally, the perturbative K\"ahler potential \eqref{eq:TreeLevKP} is independent of the axions ${\rm Re}(z^i)$, $i=1,\dots,h^{1,2}$. This leads to a discrete gauge symmetry corresponding to integer shifts of the complex structure moduli
\begin{equation}
\label{Ushift}
z^i \to z^i+n^i\,,\quad n^i\in\mathbb{Z}\, ,\quad i=1,\dots,h^{1,2}\,.
\end{equation}
The transformation of the period vector and the fluxes under these shifts is  an element of the corresponding monodromy group (which is a proper subgroup of $\text{Sp}(2h^{1,2} +2, \mathbb{Z})$):
\begin{equation}
\{\Pi,h,f\} \to M_{\{n^i\}}\{\Pi,h,f\}\,,\qquad M_{\{n^i\}}\in Sp(2h^{1,2} +2, \mathbb{Z})\,.
\end{equation}
By making use of the integer shifts in Eq.~\eqref{Ushift}, one can bring the axions to their fundamental domains ${\rm Re}(z^i)\in (-0.5,0.5]$.

Complex structure moduli stabilisation involves identifying minima of the flux-induced scalar potential \eqref{eq:scalarpotential}. In the present work, we focus on flux vacua satisfying the F-flatness conditions
\begin{subequations}
\label{eq:fflat}
\begin{align}
  D_{\tau}  W&=\frac{1}{\overline{\tau}-\tau}(f-\overline{\tau}h)^{T}\cdot \Sigma\cdot \Pi(z)=0\, , \label{eq:fflat1}\\
  D_i  W&= (f-\tau h)^{T}\cdot \Sigma\cdot\left(\partial_{i} \Pi(z)+\Pi(z)\partial_{i} K\right)=0\, . \label{eq:fflat2}
\end{align}
\end{subequations}
 We note that these conditions are equivalent to the imaginary self-duality (ISD) of the $3$-form $G_3$ \cite{Giddings:2001yu}. In terms of the flux vectors \eqref{eq:fluxdef}, this can be written as
\begin{equation}
\label{eq:ISD_complex}
    f_1-\tau\, h_1 = \overline{\vphantom{A^a}\cN\,} \cdot (f_2-\tau\,  h_2)
\end{equation}
where $\overline{\vphantom{A^a}\cN\,}$ is the (complex conjugate) gauge kinetic matrix related to the prepotential:

\begin{equation}
\label{eq:GaugeKinMatrix}
    \cN_{I J}=\overline{F}_{IJ}+2\I\, \dfrac{\text{Im}(F_{I L})X^{L} \, \text{Im}(F_{J K})X^{K}}{X^{M}\text{Im}(F_{MN})X^{N}}\kom F_{IJ}=\p_{X^{I}}\p_{X^{J}}F\, .
\end{equation}
Equivalently, by using $\tau=c_{0}+\I s$, we can write the ISD condition as
\begin{equation}
\label{eq:ISDCond_real} 
    f=\left (s\, \Sigma\cdot \cM+c_{0}\mathds{1}\right )\cdot h
\end{equation}
in terms of the real matrix
\begin{equation}
\label{eq:ISD_matrix}
    \cM = \left (\begin{array}{cc}
                     \; -\cI^{-1} &\; \cI^{-1}\cR  \\ 
                       \;\;\cR\cI^{-1} &\,  -\cI-\cR \cI^{-1}\cR\;
                       \end{array} \right )\, ,
\end{equation}
which we will call the \emph{ISD matrix}.
In the above, $\cR,\cI$ are the real and imaginary parts of the gauge kinetic matrix $\cN=\cR+\I\,\cI$ defined in \eqref{eq:GaugeKinMatrix}.

Early attempts to construct vacua solving \eqref{eq:fflat1} and \eqref{eq:fflat2} include \cite{Giryavets:2003vd,Giryavets:2004zr,DeWolfe:2004ns,Denef:2004dm,Conlon:2004ds,Eguchi:2005eh}, see also \cite{Martinez-Pedrera:2012teo,Cicoli:2013cha,Brodie:2015kza,CaboBizet:2016uzv,Blanco-Pillado:2020wjn,Blanco-Pillado:2020hbw,Lust:2024aeg} for models with $h^{1,2}\leq 3$.\footnote{An alternative strategy is to restrict to special choices of fluxes for which a subset of VEVs can be fixed analytically, see e.g. \cite{Demirtas:2019sip,Marchesano:2021gyv,Coudarchet:2022fcl}.}
The distributions of string vacua have been studied in detail (see e.g. \cite{Douglas:2003um,Ashok:2003gk,Douglas:2004zg,Denef:2004ze, Broeckel:2020fdz, Cicoli:2022chj} ) making use of the continuous flux approximation.
For any given value of $N_{\mathrm{flux}}$, the finiteness of flux vacua satisfying \eqref{eq:fflat1} and \eqref{eq:fflat2} has been proven in \cite{Grimm:2020cda,Bakker:2021uqw}.\footnote{The arguments of \cite{Grimm:2020cda,Bakker:2021uqw} actually concern self-dual classes in F-theory which extend to imaginary self-dual fluxes in the weak coupling limit for type IIB orientifolds studied in this work.}
Subsequently, the authors of \cite{Plauschinn:2023hjw} developed a constructive procedure for enumerating, at least in principle, all flux vacua in a given type IIB orientifold compactification. With this, \cite{Plauschinn:2023hjw} computationally confirmed the finiteness of F-flat vacua in a simple one-modulus case, namely an orientifold of the mirror octic with $Q_{D3}=8$.  A systematic framework for numerically constructing flux vacua was developed in \cite{Dubey:2023dvu}, making the regime $h^{1,2}\gtrsim 10$ accessible. Detailed exploration of various properties was carried out in 
\cite{Ebelt:2023clh, Chauhan:2025rdj, Chauhan:2026gid}. These studies have been in the limit of large complex
structure, the asymptotic region in moduli space where all instanton corrections to the prepotential are negligible. As discussed in the introduction, the goal of the present paper is to start probing the interior of the moduli space.

\section{A concrete two-modulus Calabi-Yau example}
\label{sec:explicit_analysis}

\subsection{Prepotential at large complex structure}

In this section, we investigate in detail the effect of worldsheet instanton corrections on moduli stabilisation. To reduce computational complexity, we focus on a Calabi-Yau orientifold with effective $h_{-}^{(1,2)} =2$. Specifically, we work at the locus of the discrete symmetry $\Gamma = {\mathbb{Z}}_{6} \times {\mathbb{Z}}_{18}$ on the degree-18 hypersurface inside $\mathbb{CP}^4_{[1,1,1,6,9]}$ see (e.g. \cite{Candelas:1994hw, Martinez-Pedrera:2012teo, Demirtas_2020} for previous studies in this setting).

As discussed in Sec.~\ref{sec:lcs}, the leading term in the prepotential is given by the intersection numbers of the mirror dual $\widetilde{X}_3$. For the effective theory of the two moduli, these are:
\begin{equation}
    \widetilde{\kappa}_{111}=9\,, \quad \widetilde{\kappa}_{112}=3\,, \quad \widetilde{\kappa}_{122}=1\,,  \quad a=\frac{1}{2} \begin{pmatrix} 9 & 3 \\ 3 & 0 \end{pmatrix}\,, \quad b=\frac{1}{4}\begin{pmatrix}
        17 \\ 6
    \end{pmatrix}\,.
\end{equation}
Now, we turn to the instanton corrections and discuss the aspects of our methodology to obtain solutions where they play a non-trivial role.
We define the truncated prepotential at degree $d$ as:
\begin{equation}
\label{prepot_truncated}
F^{(d)} = F_{\rm pert} + F^{(d)}_{\rm inst}
\end{equation}
where $F^{(d)}_{\rm inst}$ is the contribution due to all instanton terms up to degree $d$. We also denote the individual instanton correction at degree $d$ as $\Delta F^{(d)}_{\rm inst} \equiv F^{(d)}_{\rm inst}- F^{(d-1)}_{\rm inst}$. The first and second instanton corrections to the prepotential are given by:
\begin{align}
\label{eq:second_order_corrections}
    (2\I\pi)^3F^{(2)}_{\text{inst}} &=-540q_1 -3q_2-\frac{1215}{2}q_1^2 +1080q_1q_2 +\frac{45}{8}q^2_2 ,
\end{align}
with $ q_i \equiv \mathrm{e}^{2\pi \I z^i}$. A subset of the vacua constructed here belongs to a region where the instanton corrections are important enough not to be ignored. In practice, for these corrections to remain under control, we demand our F-term solutions to satisfy 

\begin{equation}
\label{eq:inst_suppression}
\frac{\left|\Delta F_{\rm inst}^{(d')}\right|}
     {\left|F^{(d)}\right|}
\le 10^{-10},
\qquad d=0\,,\,d_{\rm max}.
\end{equation}
where $d_{\rm max}$ denotes the maximum instanton degree included in the truncated prepotential used to solve the F-term equations \eqref{eq:fflat} and $d' \gg d$. We demand that instanton corrections at very high degree are suppressed. The first condition, $d =0$, ensures that the perturbative minimum already lies in a region where the instanton series is convergent. The second condition, on the non-perturbative minimum, guarantees that once instanton corrections up to degree $d_{\rm max}$ are included, the higher-degree contributions remain numerically negligible. In practice, we work with $d_{\rm max} =10$ and $d' = 100$ is chosen to make sure there are no asymptotic divergences; a detailed analysis of the convergence of these instanton corrections is presented in App.~\ref{App:convergence_region}.

A further subtlety is that a small total instanton correction may arise from an accidental internal cancellation between the two degree $d=1$ instanton components in (\ref{eq:second_order_corrections}). To exclude this possibility, we will also require each contribution to satisfy individually (see \cite{Martinez-Pedrera:2012teo}):

\begin{equation}
\label{eq:inst_disp}
\begin{aligned}
\frac{540\,e^{-2\pi\,\mathrm{Im}(z^1)}}{(2\pi)^3|F_{\rm pert}|}
\le 10^{-3}\qquad\text{and} \qquad
\frac{3\,e^{-2\pi\,\mathrm{Im}(z^2)}}{(2\pi)^3|F_{\rm pert}|}\le 10^{-3}\,.
\end{aligned}
\end{equation}

To investigate the impact of worldsheet instanton corrections on moduli stabilisation, we scan over flux quanta, restricting ourselves to the ones for which the minimum is in the large complex structure region. The region of moduli space for our scans will be:
\begin{equation}
\label{eq:tregion}
    U \subset \biggl \{\mathrm{Re}(z^i)\in (-0.5,0.5]\, ,\; \mathrm{Im}(z^i)\in [0.5,5]\, ,\;
    c_0\in (-0.5,0.5]\, ,\; s\in \left [\frac{\sqrt{3}}{2},20\right ] \biggl \}\, .
\end{equation}
under the tadpole bound $N_{\text{flux}}\leq N_{\text{max}}$ with $N_{\text{max}}=10$.  Gauge-inequivalent vacua will be obtained by
modding out by the $\text{Sp}(6,\mathbb{Z})$ and $\text{SL}(2,\mathbb{Z})$  symmetries
(see \cite{Cicoli:2022vny} for an explicit computation of the monodromy matrices
in this model). Before getting into the result of the scan, we first present some explicit examples where the instanton corrections play a non-trivial role, highlighting various interesting phenomena.

\subsection{Explicit examples of flux vacua near the boundary of LCS}
\label{sec:examples}

For most vacua in the LCS patch, the effect of instantons is negligible.
However, close to the boundary of the LCS region, there can be cases where instantons have a non-trivial effect. Here, we present some explicit examples. In the simplest cases, instantons simply induce a large shift of perturbative vacua. In more complicated examples, the effect of the instanton terms involves, instead, the multiplicity of vacua (appearance or disappearance of vacua at different orders in the instanton expansion) and monodromy shifts.

\subsubsection{Large shifts of perturbative vacua}
\label{sec:basic_examples}

The most basic phenomenon is a large displacement of a `perturbative vacuum' due to the effect of instantons. Two illustrative examples are provided in Tabs. \ref{tab:basic_1} and \ref{tab:basic_2}. The corresponding perturbative and non-perturbative vacua differ by an appreciable displacement in moduli space, despite the fact that the first instanton correction to the prepotential is roughly three orders of magnitude smaller than the perturbative contribution ($|\Delta F^{(1)}_{\rm inst}|/|F_{\rm pert}|=0.000787$ for the example presented in Tab.~\ref{tab:basic_1} and  $|\Delta F^{(1)}_{\rm inst}|/|F_{\rm pert}|= 0.000444$  for the example in Tab.~\ref{tab:basic_2}).

\begin{table}[H]
\centering
\renewcommand{\arraystretch}{1.15}
\begin{tabular}{|c|c|c|c|}
\toprule
\textbf{Modulus}
& \textbf{Pert.}
& \textbf{Non-pert.}
& \textbf{Non-pert.} \\
& \textbf{minimum}
&
\textbf{minimum ($d=1$)}
& \textbf{minimum ($d=10$)} \\
\midrule
$\langle z^1\rangle$
& $0.26696+0.88793\,i$
& $0.41459+1.05368\,i$
& $0.41550+1.05475\,i$ \\

$\langle z^2\rangle$
& $0.48664+1.61859\,i$
& $-0.01861+1.00589\,i$
& $-0.02099+1.00295\,i$ \\

$\langle \tau \rangle$
& $-0.00758+1.06273\,i$
& $-0.03951+1.05807\,i$
& $-0.03923+1.05801\,i$ \\
\bottomrule
\end{tabular}

\caption{Vacuum expectation values of the moduli for the flux configuration $f=(7,-1,0,1,0,0)$, $h=(-3,6,2,1,0,0)$ at perturbative level and after including instanton corrections at $d=1$ and $d= 10$ in (\ref{prepot_truncated}). Note the small displacement in the $(\mathrm{Im}(z^1),\mathrm{Im}(z^2))$ plane for corrections at $d=1$ and $d= 10$, $\Delta\text{Im}(z^i)=(0.00106,\,-0.00294)$.}
\label{tab:basic_1}
\end{table}

\begin{table}[H]
\centering
\renewcommand{\arraystretch}{1.15}
\begin{tabular}{|c|c|c|c|}
\toprule
\textbf{Modulus}
& \textbf{Pert.}
& \textbf{Non-pert.}
& \textbf{Non-pert.} \\
& \textbf{minimum}
&
\textbf{minimum ($d=1$)}
& \textbf{minimum ($d=10$)} \\
\midrule
$\langle z^1\rangle$
& $ -0.05577+0.95638\,i$
& $-0.11233+  1.05181\,i$
& $-0.11274+  1.051951\,i$ \\

$\langle z^2\rangle$
& $-0.35571+1.80599\,i$
& $-0.28433+ 1.40995\,i$
& $-0.28389+ 1.40924\,i$ \\

$\langle \tau \rangle$
& $-0.32613+1.50761\,i$
& $-0.44152 + 1.36571\,i$
& $0.44253+  1.36544\,i$ \\
\bottomrule
\end{tabular}
\caption{Vacuum expectation values of the moduli for the flux configuration
$f =(-1,-4,-2,\ 1,\ 0,\ 0)$, $h=(-5,-4,\ 0,\ 0,\ 0,-2)$ at perturbative level and after including instanton corrections. Again, note the small displacement in the $(\mathrm{Im}(z^1),\mathrm{Im}(z^2))$ plane for corrections at $d=1$ and $d=10$, $\Delta\text{Im}z^i=(0.00014,\,-0.00071)$.}
\label{tab:basic_2}
\end{table}

\begin{figure}[h!]
    \centering

    \begin{subfigure}[b]{0.48\linewidth}
        \centering
        \includegraphics[width=\linewidth]{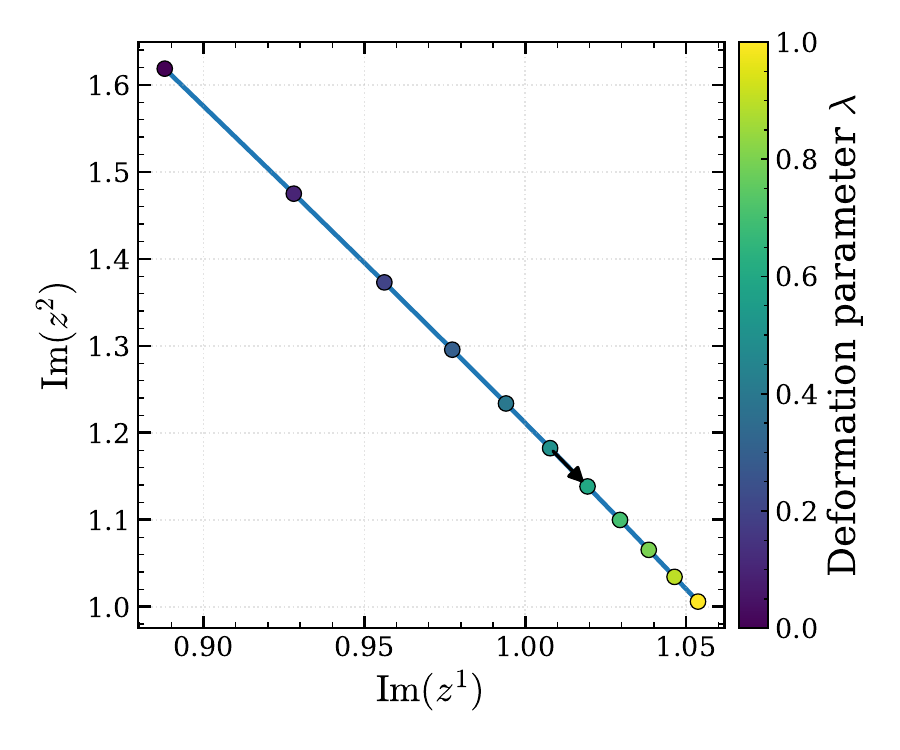}
        \label{fig:vacua_transition_a}
    \end{subfigure}
    \hfill
    \begin{subfigure}[b]{0.48\linewidth}
        \centering
        \includegraphics[width=\linewidth]{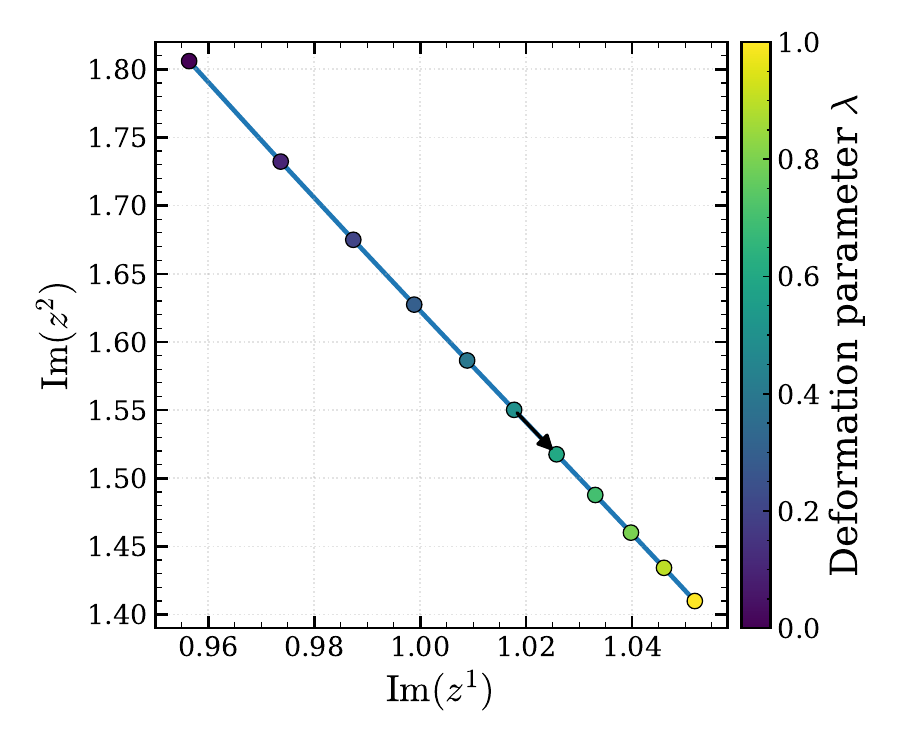}
        \label{fig:vacua_transition_b}
    \end{subfigure}
    \caption{
    Continuous transition from perturbative to non-perturbative minima for the two representative vacua in Tab.~\ref{tab:basic_1} (left) and Tab.~\ref{tab:basic_2} (right).
    The trajectories are shown in the $(\mathrm{Im}\,(z^1),\mathrm{Im}\,(z^2))$ plane. The colour of each marker
    denotes the value of the deformation parameter $\lambda\in[0,1]$.}
    \label{fig:vacua_transition}
\end{figure}

Note that the displacement of the non-perturbative vacuum between the truncations at $d=1$ and $d = 10$ is extremely small (as is seen by comparing the second
and third columns of the tables). For example in Tab. \ref{tab:basic_1}, the
shift in $(\mathrm{Im}(z^1),\mathrm{Im}(z^2))$ is  $(0.00106,\,-0.00294)$. 
Going further in the instanton expansion (in the same example), the analogous
shift between the $d= 10$ and $d= 20$ truncations is below $10^{-8}$ for both $\mathrm{Im}(z^1)$ and $\mathrm{Im}(z^2)$. Similar suppressions are also seen
for the example in Tab.~\ref{tab:basic_2}. This demonstrates that the minimum is determined by the perturbative and the first-order instantons while higher-degree instantons lead to negligible corrections. The observed displacement is a consequence of the leading instanton correction and not the result of the accumulation of many higher-order terms. A couple of comments are in order:
\begin{itemize}
\item \textbf{Non-trivial role of the degree one instanton:}  Given the hierarchy in the values of $F_{\rm pert}$
and $F^{1}_{\rm inst}$, this might seem counter-intuitive. Naively, one would expect that the effect of
$F^{1}_{\rm inst}$ can be studied in perturbation theory, and that it should cause a small shift in the minimum. To see explicitly why this is not the case, let us construct such a perturbative expansion.

Let $(\tilde{z}^i_0,\tilde{\tau}_0)$ be the perturbative canonical normalised solution satisfying $\nabla V_{\rm pert}|_{(\tilde{z}^i_0,\tilde{\tau}_0)}=0$, with $V_{\rm pert}$ denoting the scalar potential obtained without instanton corrections. Let $V_{\rm inst}$ be the scalar potential obtained by including instanton corrections up to degree $d=1$. Then we can expand both potential functions in a Taylor series around the perturbative minimum. If we keep only the first non-vanishing terms of the expansions, we can consider the following linearised displacement vector:\footnote{The validity of the truncated instanton corrections in the scalar potential $V = V_{\text{tree}} + V_{\text{inst}}$, evaluated at the shifted position $(\tilde{z}^i_{\text{inst}}, \tilde{\tau}_{\text{inst}}) = (\tilde{z}^i_0, \tilde{\tau}_0) + (\delta\tilde{z}^i_0, \delta\tilde{\tau}_0)$, requires that the displacement vector $\delta\varphi = (\delta\tilde{z}^i_0, \delta\tilde{\tau}_0)$ satisfies the condition $\delta\varphi^T M \delta\varphi + \vec{b} \cdot \delta\varphi = 0$, where $M$ is the Hessian of the perturbative potential and $\vec{b}$ is the gradient of the instanton-corrected potential evaluated at the perturbative vacuum.} 
\begin{equation}
\label{dislin}
\vec{v} = - M^{-1}|_{(\tilde{z}_0^{i},\tilde{\tau}_0)}\cdot \vec{b}\,,
\end{equation}
where $\vec{b}=\nabla V_{\rm inst}|_{(\tilde{z}^i_0,\tilde{\tau}_0)}$ is the gradient of the scalar potential with the instanton corrections (up to degree $d=1$) and $M$ is the Hessian of the perturbative scalar potential. 

The vector $\vec{v}$  provides the linear response estimate for the displacement. Note that \eqref{dislin} implies that in the case of hierarchically small perturbative masses, elements of $M^{-1}$ can be large; sizeable displacements may occur even when instanton corrections to the prepotential are small. For the two examples discussed above, we find $|\vec{v}|=11.3$  and $|\vec{v}|=2.04$, respectively. Since both values are larger than unity, perturbation theory is not a good tool to study the effect of the first instanton term on the vacuum solution.  The effect of this term should then be incorporated non-perturbatively to determine the minimum (as was done in our analysis). On the other hand, a similar analysis for higher instanton corrections shows that their effect on the minimum can be captured in perturbation theory (and hence it is small).

\item \textbf{Connection between perturbative and non-perturbative vacuum:} It is important to establish that non-perturbative minima are connected to the perturbative minima. In fact, it is known that a given flux may lead to multiple physically distinct vacua. Without such a connection, these minima could well be associated with completely different roots. To address this issue, we introduce an auxiliary deformation parameter and consider the family of prepotentials: 
\begin{equation}
\label{eq:prpotential_lambda}
F^{(1)}(\lambda)=F_{\rm pert}+\lambda\, F^{(1)}_{\rm inst}, \qquad  0\leq \lambda \leq 1.
\end{equation}

Starting from a perturbative vacuum at $\lambda=0$, we track the corresponding solution by increasing values of the parameter up to $\lambda=1$. As shown in Fig. \ref{fig:vacua_transition}, for the examples discussed, the resulting trajectories are smooth and continuous throughout the interval. This means that the displaced vacua identified in our scan are the actual non-perturbative counterpart of the perturbative solution corresponding to the same flux. 
\end{itemize}

\subsubsection{Multiple non-perturbative minima and monodromy shifts}
\label{sec:more_complicated_I}

An example exhibiting similar but more intricate behaviour than the explicit examples of Sec.~\ref{sec:examples} is presented in Tab.~\ref{tab:case_I}. At perturbative level, there is a single minimum existing inside the region $U$ in (\ref{eq:tregion}), 
while the non-perturbative analysis yields two distinct minima with instanton corrections under control. Unlike the examples of Sec. \ref{sec:examples}, the additional non-perturbative solution experiences non-negligible shifts after including higher-degree instanton corrections before eventually stabilising\footnote{Here, the term \textit{stabilisation} refers to the absence of significant further shift in the moduli under successive instantons inclusion, although the F-terms are solved at every successive order.}.

\begin{table}[h!]
\centering
\renewcommand{\arraystretch}{1.15}

\begin{tabular}{cccc}
\toprule
\textbf{Modulus}
&
\textbf{Pert.}
&
\textbf{Non-pert.}
&
\textbf{Non-pert.}\\
&
\textbf{minimum}
&
\textbf{minimum I ($d=1$)}
&
\textbf{minimum I ($d=10$)}\\
\midrule

$\langle z^1\rangle$
& $-0.479908+1.036205\,i$
& $-0.497424 + 0.958081\,i$ 
& $-0.496982+0.960373\,i$\\

$\langle z^2\rangle$
& $-0.404498+0.757402\,i$
& $-0.350100+ 0.901698\,i$ 
& $-0.353310+0.897701\,i$\\

$\langle\tau\rangle$
& $0.498853+1.429942\,i$
& $0.404802+1.474379\, i$ 
& $0.407117+1.473142\,i$\\
\midrule
&
&
\textbf{minimum II ($d=1$)}
&
\textbf{minimum II ($d=10$)}\\
\midrule

$\langle z^1\rangle$
&---
& $-0.371980+ 0.638758\,i$ 
& $-0.334308+0.605470\,i$\\

$\langle z^2\rangle$
&---
& $-0.434522+1.992250\,i$ 
& $-0.486847+2.224954\,i$\\

$\langle\tau\rangle$
&---
& $-0.020261+ 1.908438\,i$
& $-0.054286+2.036671\,i$\\
\bottomrule
\end{tabular}
\caption{ Multiple non-perturbative minima for the flux configuration $f=(12,16,3,-1,0,4)$, $h=(3,5,0,0,0,2)$. The perturbative potential admits a single minimum inside $U$, while the inclusion of instanton corrections gives rise to two distinct non-perturbative minima. The values shown correspond to the truncations at $d=1$ and $d=10$ in (\ref{prepot_truncated}).}
\label{tab:case_I}
\end{table}

\begin{figure}[h!]
    \centering
    \begin{subfigure}{0.495\linewidth}
        \centering
        \includegraphics[width=\linewidth]{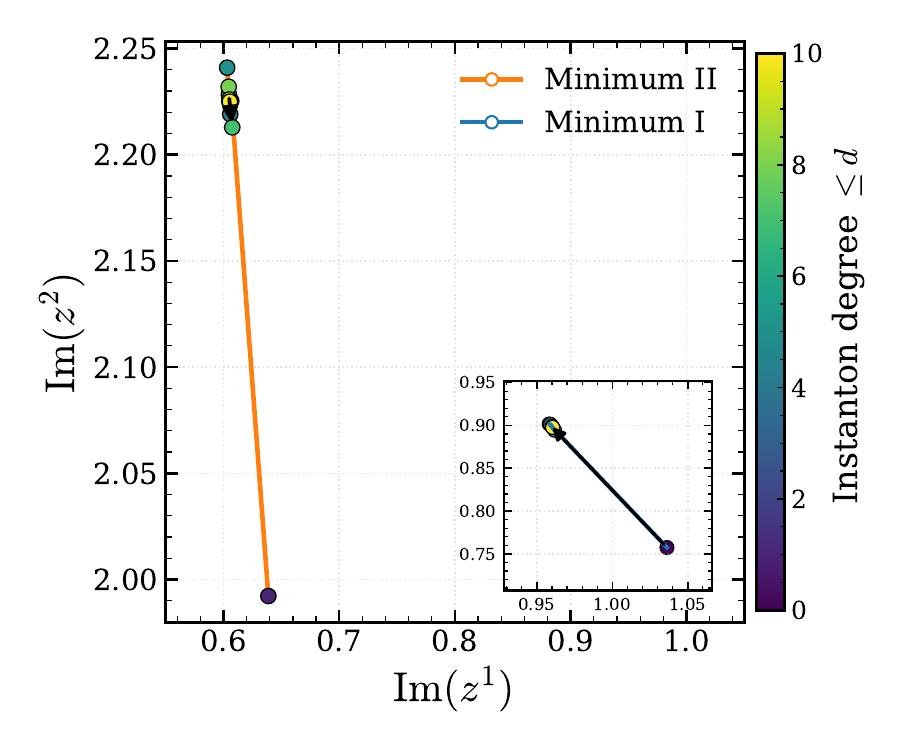}
        \label{fig:inst_track}
    \end{subfigure}
    \hfill
    \begin{subfigure}{0.49\linewidth}
        \centering
        \includegraphics[width=\linewidth]{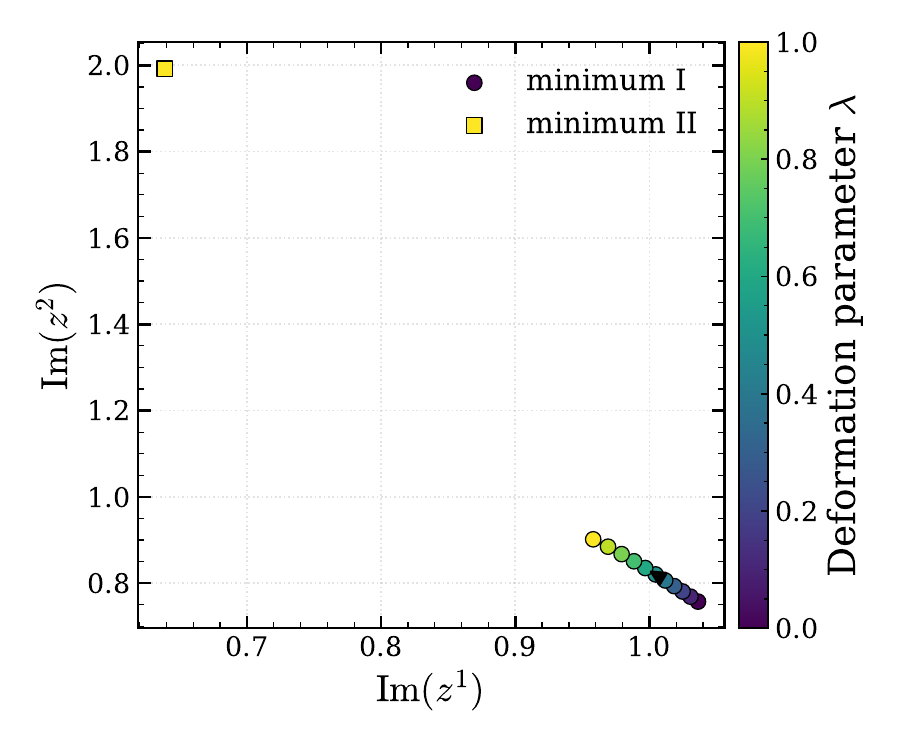}
        \label{fig:deform_track}
    \end{subfigure}
\caption{Evolution of the different minima in Tab.~\ref{tab:case_I} 
    under deformations. \textit{Left}: Trajectories of the minima under successive instanton inclusion in the prepotential. The orange minimum enters $U$ at degree $d=1$ and shifts further before "stabilising" up to degree $d= 10$. \textit{Right}: Evolution of the same minima under a continuous deformation of the prepotential, with the emergence of an additional minimum at degree 1 shown as an isolated yellow point.}
    \label{fig:tracks_case_I}
\end{figure}

The evolution of the two branches under the continuous prepotential deformation, as described in Sec.~\ref{sec:explicit_analysis}, is shown in Fig.~\ref{fig:tracks_case_I}. As illustrated in the left panel, the minimum shown in blue continuously deforms from the perturbative solution and becomes essentially stable already after the inclusion of the first instanton corrections, with negligible changes at higher degrees. By contrast, the second branch, shown in orange, has no perturbative counterpart inside $U$: it first appears upon including the leading instanton correction and undergoes a further shift at degree two before stabilising. This behaviour is corroborated by the right panel, where the corresponding branch emerges only at $\lambda=1$ in the interpolating prepotential (\ref{eq:prpotential_lambda}), demonstrating that this vacuum is generated inside $U$ by the instanton correction.

A closer inspection reveals that the missing perturbative minimum in Tab.~\ref{tab:case_I} does in fact exist, but lies outside the fundamental region $U$: $ \langle z^1 \rangle =0.245515+ 0.512225\, i, \langle z^2 \rangle =-1.768403+2.827649\, i, \langle \tau \rangle = 2.510555+  2.732775\, i$. However, after performing the appropriate $\text{Sp}(6,\mathbb{Z})$ monodromy together with the $\text{SL}(2,\mathbb{Z})$ duality transformation, this perturbative solution can be mapped back into the fundamental domain with fluxes and moduli vevs given by:
\begin{equation}
    f = (-9, 4, 3, -1, 0, -4) \,, \qquad 
    h = (3, 1, 0, 0, 0, 2) \,,
\end{equation}
\begin{align}
    \langle z^1 \rangle &= \phantom{-}0.245515 + 0.512225 \, i \,, \nonumber \\
    \langle z^2 \rangle &= \phantom{-}0.231596 + 2.827649 \, i \,, \\
    \langle \tau \rangle &= -0.489444 + 2.732775 \, i \,. \nonumber
\end{align}
In fact, this mapped solution is found to evolve precisely into the second non-perturbative branch since the non-perturbative solution for this transformed flux at $d=1$ and $d=10$ matches the non-perturbative minimum II listed in Tab.~\ref{tab:case_I} after  $\text{Sp}(6,\mathbb{Z})$ and $\text{SL}(2,\mathbb{Z})$ transformations. The agreement between the transformed perturbative vacuum and the non-perturbative solution therefore provides a consistency check of the monodromy identification. 

\subsubsection{Multiple perturbative minima and exit from LCS}
\label{sec:more_complicated_II}

There are cases where not all multiple perturbative minima occurring inside $U$ associated with a flux configuration survive as non-perturbative minima. One such example is shown in Tab.~\ref{tab:case_II}. At perturbative level, the flux configuration admits two distinct solutions inside $U$. However, after including instanton corrections, only the branch of minimum II remains inside the LCS patch. The second perturbative branch (minimum I) is driven towards the boundary of the region of convergence and eventually moves beyond it. This implies that out of the multiple perturbative minima, the ones close to the boundary of the region of convergence\footnote{See App.~\ref{App:convergence_region} for details on the convergence region.} of the instanton expansion may destabilise (under instantons), leaving a single unique non-perturbative minimum. In this sense, instanton effects may lift the multiplicity present at the perturbative level. This is opposite to the behaviour presented in Sec. \ref{sec:more_complicated_I}. 

\begin{table}[H]
\centering
\renewcommand{\arraystretch}{1.15}

\begin{tabular}{cccc}
\toprule
\textbf{Modulus}
&
\multicolumn{3}{c}{\textbf{Minima}}\\
\cmidrule(lr){2-4}
&
\textbf{Pert. I}
&
\textbf{Pert. II}
&
\textbf{Non-pert.} ($d =1$)\\
\midrule
$\langle z^1\rangle$
& $-0.231694+0.900115\,i$
& $-0.385852+1.141803\,i$
& $-0.378369+1.125654\,i$ \\

$\langle z^2\rangle$
& $-0.456397+3.085719\,i$
& $-0.414998+1.702908\,i$
& $-0.430900+1.774539\,i$ \\

$\langle\tau\rangle$
& $0.289944+3.122602\,i$
& $-0.108075+2.360777\,i$
& $-0.137013+2.399982\,i$ \\
\bottomrule
\end{tabular}

\caption{Multiple perturbative minima for the flux configuration
$f=(7,8,2,-1,0,2)$, $h=(4,3,0,0,0,1)$. The perturbative potential admits two distinct minima, which upon instanton correction lead to a single non-perturbative minimum. The vacuum position remains unchanged upon incorporating instanton corrections beyond degree $d=1$. The shift in the Im$(z_{\rm inst}^i)$-plane between $d=1$ and up to $d=10$ instanton is $(0.000118,-0.000588)$.}
\label{tab:case_II}
\end{table}

A similar example of a flux configuration that admits a perturbative solution which disappears from $U$ after the inclusion of instantons is presented in Tab.~\ref{tab:tree_example}.

\begin{table}[h!]
\centering
\renewcommand{\arraystretch}{1.15}
\begin{tabular}{ccc}
\toprule
\textbf{Modulus}
&
\textbf{Pert. minimum}
&
\textbf{Non-pert. minimum}\\
\midrule
$\langle z^1\rangle$
&
$0.347104+0.705096\,i$
&
--- \\
$\langle z^2\rangle$
&
$0.471862+2.401130\,i$
&
--- \\
$\langle\tau\rangle$
&
$0.133949+8.233088\,i$
&
--- \\
\bottomrule
\end{tabular}
\caption{Instanton-destabilised minimum for the flux configuration
$f=(-19,8,6,-2,0,-6)$,
$h=(2,0,0,0,0,1)$. The perturbative potential admits a minimum, but no corresponding non-perturbative minimum is found inside $U$ upon the inclusion of worldsheet instanton corrections.}
\label{tab:tree_example}
\end{table}

We emphasise that in both examples, the disappearing perturbative minimum in Tabs. \ref{tab:case_II} and \ref{tab:tree_example} leaves not only the fundamental domain $U$ but also the region of convergence in the Im${(z^i)}$-plane. In other words, these examples illustrate that instantons may destabilise perturbative minima located close to the boundary of the LCS patch.

\subsubsection{Instanton-induced minima at higher order} 
\label{sec:more_complicated_III}

Instanton corrections may also generate vacua that are absent at perturbative level. The example shown in Tab.~\ref{tab:case_III} illustrates this possibility. For this flux choice, no minimum exists within\footnote{We did not find any perturbative minimum connected to the non-perturbative minimum, even outside the fundamental domain for the monodromy $\rm{Sp}(6, \mathbb{Z})$.} $U$ at perturbative level, whereas a stable solution appears in $U$ once instantons up to degree $d\leq2$ are included.
This is a genuine case of instanton-induced vacuum, indicating the incompleteness of the classical large complex structure vacua near the boundary.  

\begin{table}[H]
\centering
\renewcommand{\arraystretch}{1.15}

\begin{tabular}{ccc}
\toprule
\textbf{Modulus}
&
\textbf{Pert. minimum}
&
\textbf{Non-pert. minimum} $(d=2)$\\
\midrule
&
\multicolumn{1}{c}{No vacuum in $U$}
& \\[-1.5ex]
$\langle z^1\rangle$
&
---
&
$0.460237+0.546043\,i$ \\
$\langle z^2\rangle$
&
---
&
$0.345072+3.689011\,i$ \\
$\langle\tau\rangle$
&
---
&
$-0.442060+9.478178\,i$ \\
\bottomrule
\end{tabular}
\caption {Instanton-induced minimum for the flux configuration
$f=(-27,11,8,-1,0,-10)$, $h=(2,0,0,0,0,1)$.
No perturbative vacuum is found in the search region $U$, while a non-perturbative vacuum first appears at instanton degree $d = 2$ with a shift $\Delta \text{Im}z^i=(-0.00210, 0.007946)$ between $d=2$ and $d= 10$.}
\label{tab:case_III}
\end{table}
Another example is presented in Tab.~\ref{tab:case_IV}. For this flux configuration, the perturbative theory admits no solution inside $U$, whereas two distinct non-perturbative minima appear upon including the leading instanton corrections. Both minima experience negligible shifts due to higher instantons. Also in this case, no perturbative minimum is found even after relaxing the condition to lie inside the fundamental domain.

\begin{table}[h!]
\centering
\renewcommand{\arraystretch}{1.15}

\begin{tabular}{ccc}
\toprule
\textbf{Modulus}
&
\textbf{Pert. minimum}
&
\textbf{Non-pert. minimum} $(d= 1)$\\
\midrule
&
\multicolumn{1}{c}{No vacuum in $U$ }
&
\textbf{I}\\
\cmidrule(lr){3-3}
$\langle z^1\rangle$
&
---
&
$0.424227+1.123743\,i$ \\
$\langle z^2\rangle$
&
---
&
$-0.395179+1.240314\,i$ \\
$\langle\tau\rangle$
&
---
&
$0.435179+1.190339\,i$ \\
\midrule
&
&
\textbf{II}\\
\cmidrule(lr){3-3}
$\langle z^1\rangle$
&
---
&
$0.289284 + 0.887175\,i$ \\
$\langle z^2\rangle$
&
---
&
$0.226593 +
         1.550260\,i$ \\
$\langle\tau\rangle$
&
---
&
$0.065203 + 1.106972\,i$ \\
\bottomrule
\end{tabular}
\caption{Instanton-induced minima for the flux configuration
$f=(-6,-5,0,-1,0,-1)$, $h=(1,-4,-2,-1,-1,2)$.
No perturbative minimum is found in the search region $U$, while two distinct non-perturbative minima are generated upon the inclusion of worldsheet instanton corrections with shifts $\Delta\text{Im}=(0.00020,-0.00026), (0.00432,-0.00580)$, respectively, between $d=1$ and $d=10$.}
\label{tab:case_IV}
\end{table}

\begin{figure}[h!]
    \centering
    \begin{subfigure}[b]{0.48\linewidth}
        \centering
        \includegraphics[width=\linewidth]{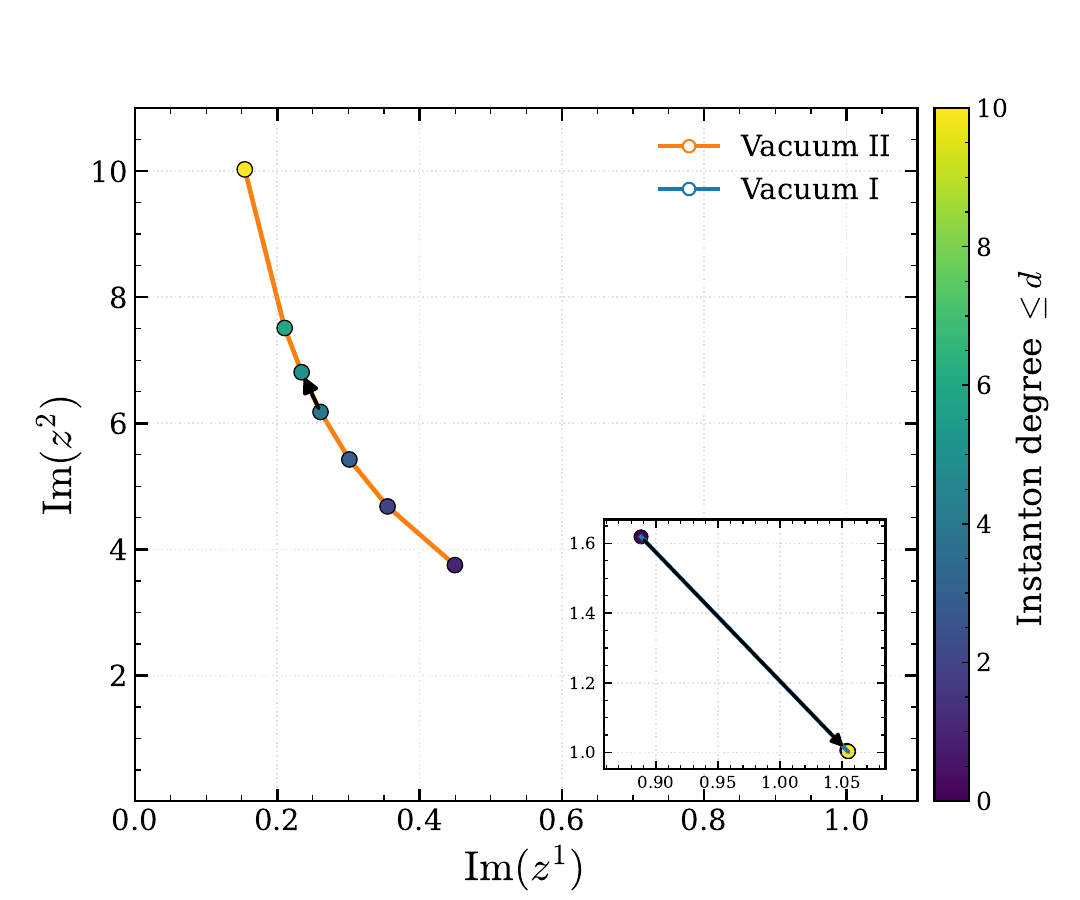}
    \end{subfigure}
    \hfill
    \begin{subfigure}[b]{0.48\linewidth}
        \centering
        \includegraphics[width=\linewidth]{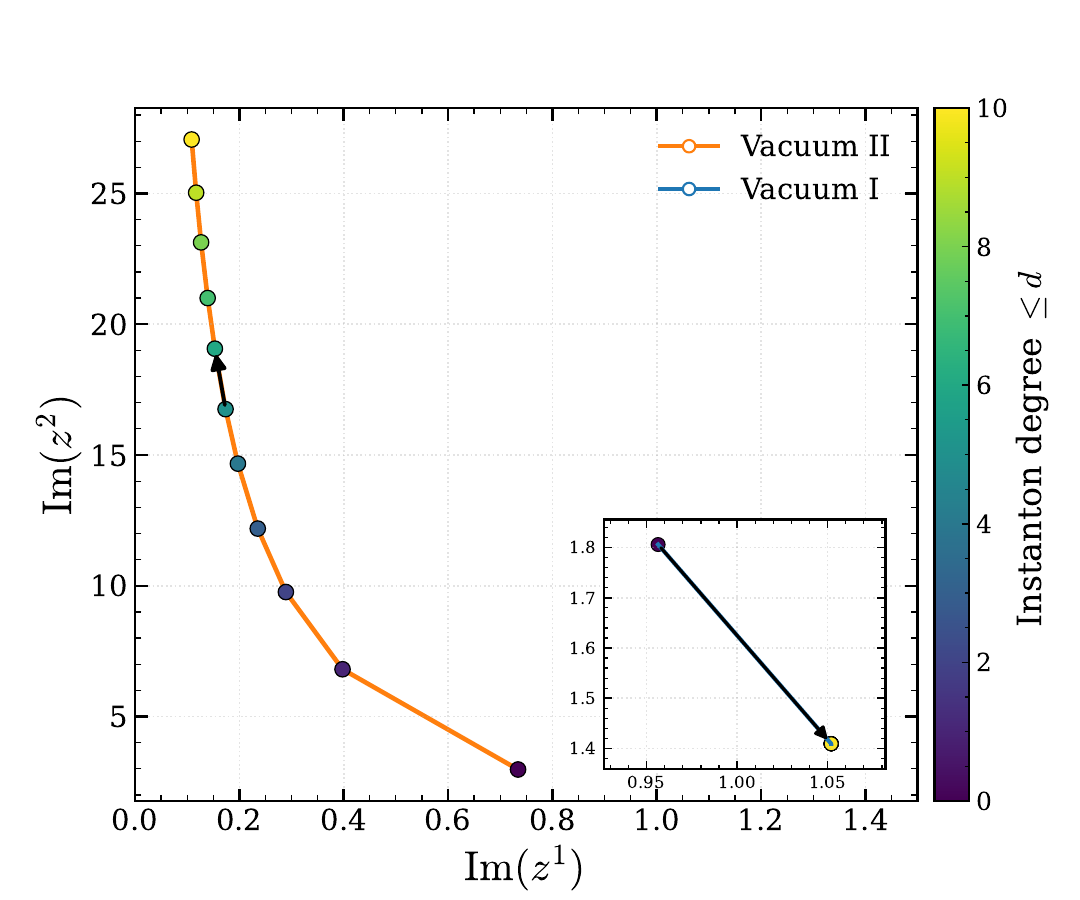}
    \end{subfigure}
    \caption{Evolution of the additional non-perturbative minimum under successive instanton corrections for the flux configurations listed in Tab.~\ref{tab:basic_1} (left) and Tab.~\ref{tab:basic_2} (right). The unstable branches (orange) persist only up to a finite instanton degree, whereas the stable ones (blue) survive throughout the instanton expansion and converge to the fully non-perturbative minima reported in Tab.~\ref{tab:basic_1} and Tab.~\ref{tab:basic_2}. The colour of each marker indicates the maximum instanton degree included in the truncation.}
    \label{fig:basic_unstable}
\end{figure}

Lastly, it turns out that there are additional minima for the explicit example presented in Tabs. \ref{tab:basic_1} and \ref{tab:basic_2}. Unlike the stable branches discussed in Sec. \ref{sec:basic_examples}, these solutions, appearing at leading instanton correction, undergo large displacements as successively higher-degree instanton are included and eventually disappear for $d > 10$. Such behaviour is characteristic of an unstable branch generated by an insufficient early truncation of the instanton expansion. The trajectories of these roots (orange line), along with the stable counterparts (blue line) with inclusion of successive instanton corrections, are shown in Fig.~\ref{fig:basic_unstable}.  

These examples together illustrate that worldsheet instanton corrections can affect the vacuum structure in various qualitative ways. In the next section, we investigate how frequently the basic non-perturbative behaviour occurs within the full ensemble of flux vacua we have collected in our scan. 

\section{Statistical analysis of flux vacua near the boundary of LCS}
\label{sec:pvsnp}

The explicit examples of the previous sections establish that worldsheet instantons can significantly affect the perturbative solutions. We now address the complementary question of whether these examples represent isolated cases or instead constitute a statistically significant feature of the landscape. We perform a scan for non-perturbative vacua in the LCS and restrict the search to a bounded region of the moduli space with $1 \leq \text{Im}(z^i) \leq 5$ and $N_{\text{max}}\leq 10$. 

\begin{figure}[h]
\centering
\includegraphics[width=1\linewidth]{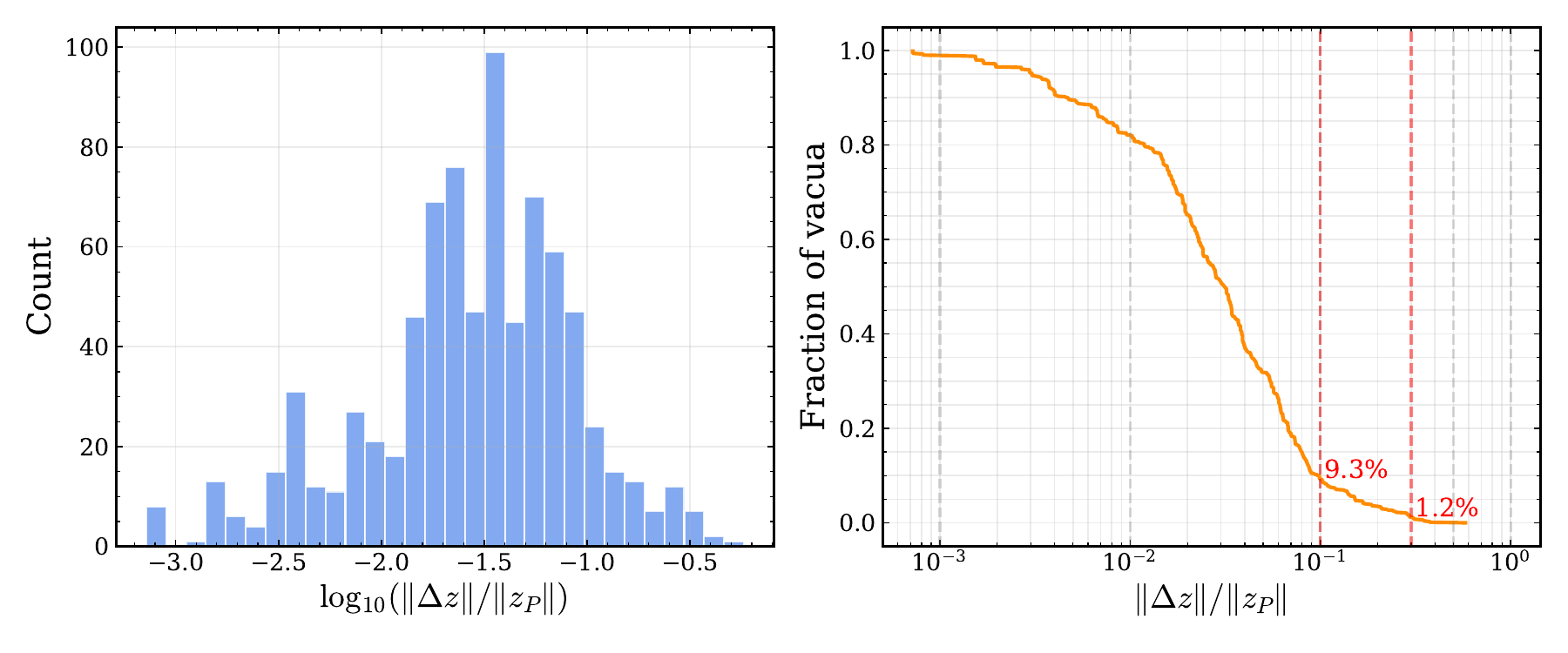}
\caption{\textit{Left}: Distribution of the relative vacuum displacement on a logarithmic scale between perturbative and non-perturbative vacua. While most vacua are only weakly affected by instanton corrections, a statistically significant tail exhibits sizeable shifts. \textit{Right}: Survival function quantifying the fraction of vacua whose relative displacement exceeds a given threshold $||\Delta z/z ||$. The analysis is restricted to vacua satisfying $\min\{\mathrm{Im}(z_1),\mathrm{Im}(z_2)\}<1$, where worldsheet instantons are expected to have the largest impact. The fractions of vacua with $||\Delta z/z ||\geq 0.1, 0.3$ are $\simeq 9.3 \%, 1.2 \%$, respectively.}
    \label{fig:vacua_hist}
\end{figure}

For each instanton corrected vacuum, we identify its perturbative counterpart and compute the relative displacement. 
Since worldsheet instantons are expected to become relevant only near the boundary of the large complex structure patch, we can focus the statistical analysis on the subset of solutions with $\text{min}\{\text{Im}(z^i)\}<1$. In Fig. \ref{fig:vacua_hist} we summarise the resulting distribution. As expected, for the majority of the flux configurations, the displacement of the non-perturbative vacuum compared to the perturbative one is extremely small. These vacua remain essentially unchanged after instanton corrections are included, confirming the usual expectation that exponentially suppressed corrections should produce only minor modifications of the perturbatively stabilised solutions. However, as shown in Fig. \ref{fig:vacua_hist}, the complete ensemble exhibits a more interesting structure. While the first instanton correction typically remains about three orders of magnitude smaller than the perturbative component, a non-negligible subset of vacua displays significantly larger displacements than would be expected from the size of the instanton corrections. In particular, we find examples for which the relative shift in the canonical normalised moduli is of order unity. As quantified in the right panel of the figure, the fraction of vacua whose relative displacement is above the threshold $0.1$ turns out to be $\simeq 9.3 \%$.

\begin{figure}[h]
\centering
\includegraphics[width=1\linewidth]{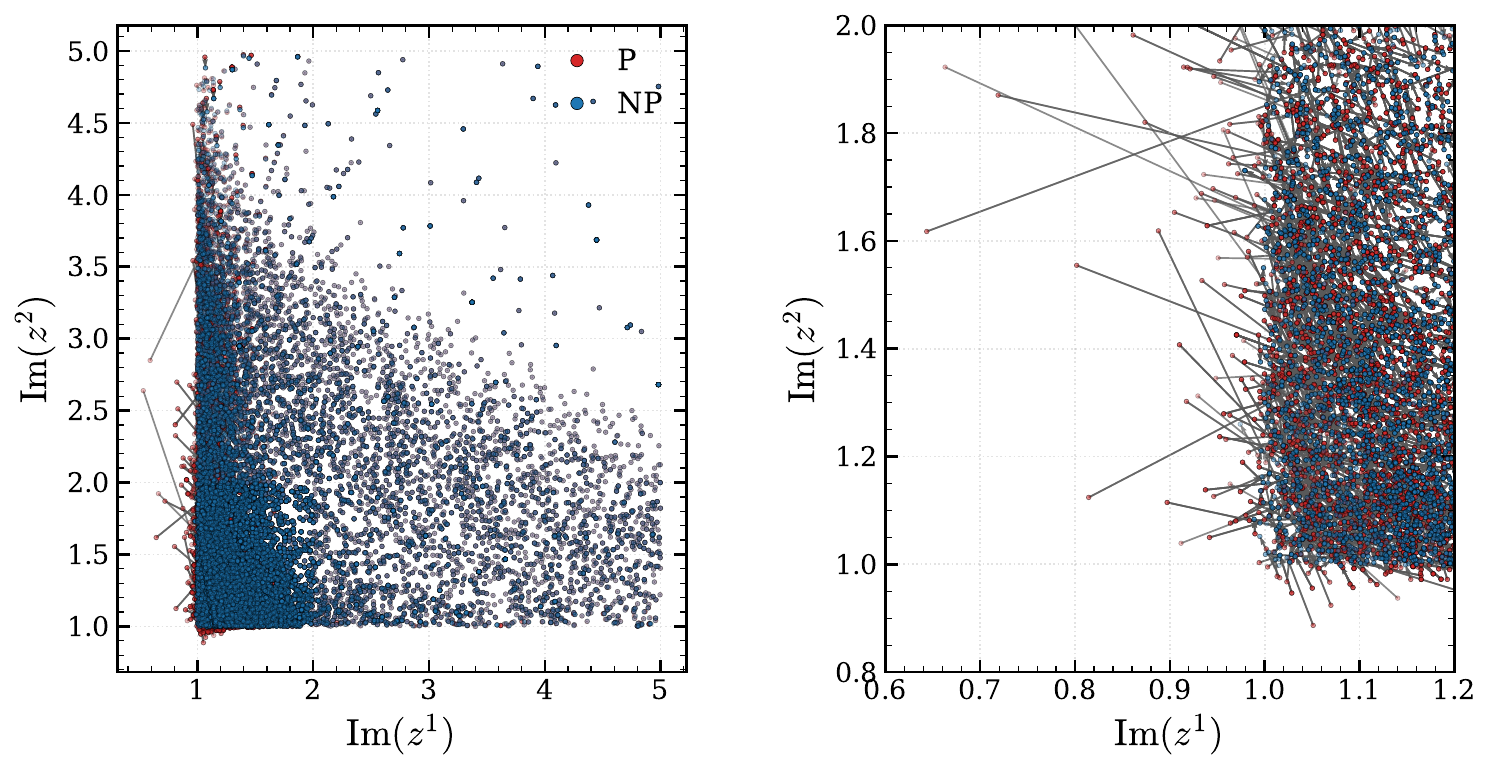}
\caption{Displacement of perturbative flux vacua induced by worldsheet instanton corrections in the Im$(z_1)$, Im$(z_2)$ plane. Perturbative and non-perturbative vacua are shown in red and blue, respectively. The arrows connect each perturbative vacuum to its continuously connected non-perturbative counterpart. The right panel is a zoomed view of the region with the highest density of vacua, where the largest displacements are observed close to the boundary of the LCS patch.}
\label{fig:vacua_shifts}
\end{figure}

The spatial distribution of vacuum displacements in moduli space is shown in Fig. \ref{fig:vacua_shifts}. Large shifts are not uniformly distributed throughout moduli space, but instead they are particularly pronounced and clustered close to the lower boundary of the scanned large complex structure region, where the imaginary parts of the complex structure moduli are of order unity. In contrast, vacua located deeper in the LCS regime exhibit negligible shifts and remain nearly indistinguishable from their perturbative roots. 
This observed behaviour is consistent with the expectation that the first instanton correction becomes phenomenologically relevant precisely where it is less exponentially suppressed (see App. \ref{App:convergence_region}), while higher instantons remain mostly negligible.

\begin{figure}[H]
\centering
\includegraphics[width=0.7\linewidth]{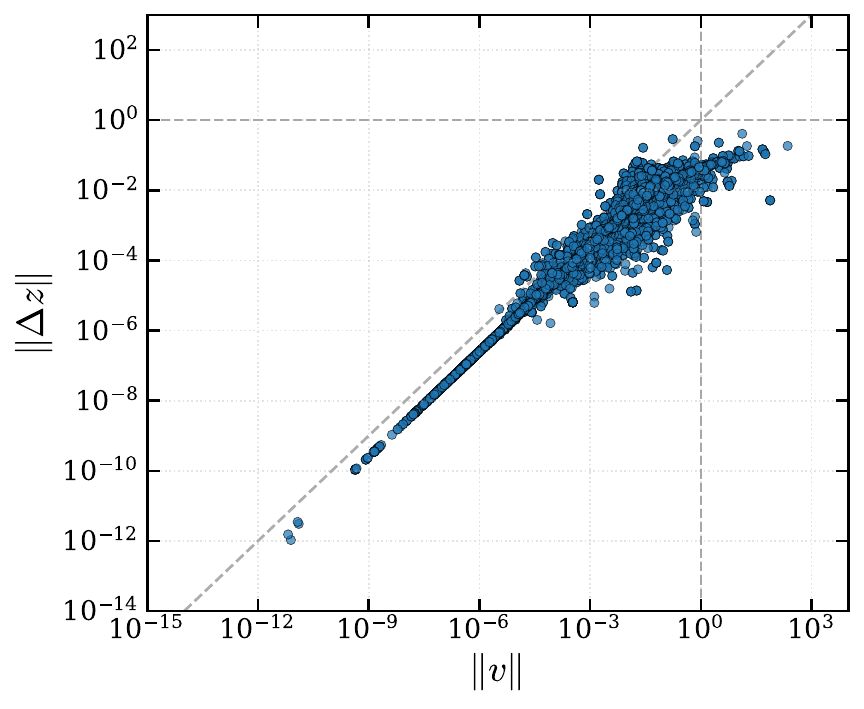}
\caption{Comparison between the norm of the linearised displacement vector $||\vec{v|}|=||- M^{-1}\cdot \vec{b}||$ and the corresponding actual displacements of the canonical moduli,  $||\Delta z ||\equiv||(\tilde{z}^i_{\rm inst},\tilde{\tau}_{\rm inst})-(\tilde{z}^i_0,\tilde{\tau}_0)||$. The dashed line corresponds to perfect agreement between the linearised estimate and the full non-perturbative solution. The approximately linear correlation demonstrates that non-perturbative effects become relevant for large $||v||$.}
\label{fig:v_vecvsdx}
\end{figure}

The perturbative interpretation developed in the representative examples can be applied to the entire ensemble. For every perturbative vacuum we evaluate the vector $\vec{v} = - M^{-1}|_{(\tilde{z}_0^{i},\tilde{\tau}_0)}\cdot \vec{b}$, defined in Sec. \ref{sec:examples}, and compare its norm with the actual canonical displacement between the corresponding perturbative and the non-perturbative vacua, $||(\tilde{z}^i_{\rm inst},\tilde{\tau}_{\rm inst})-(\tilde{z}^i_0,\tilde{\tau}_0)||$. In Fig. \ref{fig:v_vecvsdx} we find an overall approximately linear correlation over several orders of magnitude. The deviations from the expectation are mostly below the dashed line, confirming the control over the non-perturbative solutions. This provides strong evidence that the observed large vacuum shifts are explained in terms of large-norm $\vec{v}$ vectors for which we have the breakdown of the perturbative approximation.

The statistical analysis conducted here focuses only on perturbative vacua that admit a continuously connected non-perturbative counterpart, as for the examples in Sec. \ref{sec:basic_examples}. A systematic statistical analysis for the more intricate behaviour involving multiplicities, as discussed in Sec. \ref{sec:more_complicated_I}, is left for future work.

\section{Conclusions}
\label{sec:summary}

In this study, we have probed IIB flux vacua near the boundary of the large complex structure region, i.e. vacua in the large complex structure patch with the first instanton terms playing a non-trivial role. We have seen various novel features. The simplest are vacua where the minimum is determined by the interplay of the perturbative terms and the first instanton correction (with all other terms negligible); such a vacuum
can be thought of as arising from a large displacement of a perturbative vacuum caused by the instantons. We have also observed more subtle phenomena involving the multiplicities of vacua and monodromy shifts. Our statistical analysis shows that non-trivial effects from instantons are quite generic.

We also used systematic procedures for testing the control over the instanton correction at higher and higher degrees; the necessity of these tests becomes more evident close to the boundary of the LCS patch, and non-trivial effects from instantons can occur quite generally. 
      
Overall, our study makes it clear that we should expect novel phenomena and a rich structure as we probe string theory away from the asymptotics of moduli spaces. It also opens up several interesting avenues for future study. So far, we have not discovered solutions where a large number of instanton corrections play a non-trivial role. One can, however, expect such solutions to be present a bit closer to the boundary of the LCS patch. It will be interesting to find such solutions explicitly. Another interesting direction is to study K\"ahler moduli stabilisation in the context of the solutions presented in this article. On statistical considerations, it is important to understand how the number of vacua where the instantons play a non-trivial role scales as the number of complex structure moduli is increased.
  
\section*{Acknowledgments}

We thank Sven Krippendorf, Ratul Mahanta and Andreas Schachner for fruitful discussions. AM would like to thank the Leinweber Institute for Theoretical Physics for supporting his sabbatical visit at the University of Michigan, Ann Arbor. The work of PP is supported by a NYUAD research grant. We want to thank the creators of the \texttt{CYTools} \cite{Demirtas:2022hqf,Demirtas:2023als} and \texttt{JAXVacua} \cite{Dubey:2023dvu}.

\appendix

\section{Gopakumar-Vafa invariants}
\label{app:gv_invariants}

We now express the relationship between the default basis vectors for the Mori cone used in the \texttt{CYTools} package and the  K\"{a}hler cone basis. We define the K\"ahler cone as the positive first quadrant Im$({z^i})\geq 0$, such that the instanton corrections are exponentially suppressed in the LCS, i.e. as Im$({z^i})\rightarrow \infty$. This requires the default basis of the Mori cone
\begin{equation}
\quad
C_1 = (1,-3), \quad C_2 = (0,1),    
\end{equation}
to be rotated so that the dual K\"ahler cone is generated by the canonical basis of the positive quadrant. The matrix for the required operation is:
\begin{equation}
   A =
\begin{pmatrix}
 3 & 1\\
 1 & 0
\end{pmatrix}. 
\end{equation}

We generated the genus-zero GV invariants $n_{d_1,d_2}$ 
for the Calabi--Yau threefold $\mathbb{CP}^4[1,1,1,6,9]$ up to total degree $d_1 + d_2 \leq 300$. 
These invariants enter the instanton expansion of the prepotential as 
\begin{equation}
{F}_{\mathrm{inst}}(q_1, q_2) = \sum_{(d_1, d_2) \neq (0,0)} n_{d_1,d_2} \, \text{Li}_3(q_1^{d_1} q_2^{d_2}) = \sum_{(d_1, d_2) \neq (0,0)} \sum_{k=1}^{\infty} n_{d_1,d_2} \frac{q_1^{k d_1} q_2^{k d_2}}{k^3}
\end{equation}
where, $q_i = \exp(2\pi i z^i)$.
By rearranging the summation indices, we can express the prepotential directly in terms of the Gromov-Witten invariants $N_{d_1,d_2}$:
\begin{equation}
{F}_{\mathrm{inst}}(q_1, q_2) = \sum_{(d_1, d_2) \neq (0,0)} N_{d_1,d_2} \, q_1^{d_1} q_2^{d_2}
\end{equation}
where the relationship between the two sets of invariants is given by the multicovering formula:
\begin{equation}
N_{d_1,d_2} = \sum_{k | \gcd(d_1, d_2)} \frac{n_{d_1/k, \, d_2/k}}{k^3} \ .
\end{equation}
The resulting instanton contribution to the prepotential up to degree $10$ takes the form:

\begingroup
\footnotesize
\begin{align}
(2\pi \mathrm{i})^3 F^{(10)}_{\text{inst}} &= -540 q_1 - 3 q_2 \nonumber \\
&\quad - \frac{1215}{2} q_1^2 + 1080 q_1 q_2 + \frac{45}{8} q_2^2 \nonumber \\
&\quad - 560 q_1^3 - 143370 q_1^2 q_2 - 2700 q_1 q_2^2 - \frac{244}{9} q_2^3 \nonumber \\
&\quad - \frac{39420}{64} q_1^4 - 204071184 q_1^3 q_2 + 574695 q_1^2 q_2^2 + 17280 q_1 q_2^3 + \frac{12333}{64} q_2^4 \nonumber \\
&\quad - \frac{68040}{125} q_1^5 - 21772947555 q_1^4 q_2 - 74810520 q_1^3 q_2^2 - 5051970 q_1^2 q_2^3 - 154440 q_1 q_2^4 - \frac{211878}{125} q_2^5 \nonumber \\
&\quad - 630 q_1^6 - 1076518252152 q_1^5 q_2 + 49933131480 q_1^4 q_2^2 + 913383040 q_1^3 q_2^3 \nonumber \\
&\quad + 57882060 q_1^2 q_2^4 + 1640520 q_1 q_2^5 + \frac{102365}{6} q_2^6 \nonumber \\
&\quad - 540 q_1^7 - 33381348217290 q_1^6 q_2 - 7772494870800 q_1^5 q_2^2 - 224108858700 q_1^4 q_2^3 \nonumber \\
&\quad - 13593850920 q_1^3 q_2^4 - 751684050 q_1^2 q_2^5 - 19369800 q_1 q_2^6 - \frac{64639725}{343} q_2^7 \nonumber \\
&\quad - \frac{39420}{64} q_1^8 - 746807207168880 q_1^7 q_2 - 31128163315047072 q_1^6 q_2^2 + 42712135606368 q_1^5 q_2^3 \nonumber \\
&\quad + 2953943406210 q_1^4 q_2^4 + 218032516800 q_1^3 q_2^5 + 10500261120 q_1^2 q_2^6 + 245635200 q_1 q_2^7 + \frac{1140830253}{512} q_2^8 \nonumber \\
&\quad - 560 q_1^9 - 13066023094376184 q_1^8 q_2 - 8211715737128556480 q_1^7 q_2^2 + 16612333123572659520 q_1^6 q_2^3 \nonumber \\
&\quad - 603778002921828 q_1^5 q_2^4 - 51350781706785 q_1^4 q_2^5 - 3630383423100 q_1^3 q_2^6 \nonumber \\
&\quad - 153827405370 q_1^2 q_2^7 - 3279587940 q_1 q_2^8 - \frac{6742982701}{243} q_2^9 \nonumber \\
&\quad - \frac{68040}{125} q_{1}^{10} - 188271614342884440 q_1^9 q_2 - 1028507105335081958010 q_1^8 q_2^2 \nonumber \\
&\quad + 16612333123572659520 q_1^7 q_2^3 + 90433961251273800 q_1^6 q_2^4 + 11035406089270080 q_1^5 q_2^5 \nonumber \\
&\quad + 967920854160960 q_1^4 q_2^6 + 61789428573120 q_1^3 q_2^7 + 2330291414880 q_1^2 q_2^8 \nonumber \\
&\quad + 45523225800 q_1 q_2^9 + 360012150 q_{2}^{10} \ .
\end{align}
\endgroup

\section {Region of convergence of the instanton expansion}
\label{App:convergence_region}

In Secs. \ref{sec:explicit_analysis} and \ref{sec:pvsnp} we saw that the leading instanton corrections can induce large displacements of the perturbative flux vacua and that this behaviour is a statistically significant feature of our ensemble of solutions. In Sec. \ref{sec:pvsnp}, in particular, we saw that this behaviour is a feature of a specific region of the moduli space which, as we will show in this appendix, is the result of the convergence of the instanton expansion.

We already stressed multiple times that in all our solutions the instanton expansion is under control. A large displacement indicates that the leading instanton correction competes locally with the perturbative potential, but it does not imply that higher-degree instantons become important. In this appendix, we quantify the convergence of the prepotential expansion and identify the region of the moduli space where the truncation adopted throughout this work is quantitatively reliable and the leading instanton correction is large enough to compete with the locally suppressed perturbative term.

We compute all instanton corrections for $ \mathbb{CP}^4_{[1,1,1,6,9]} $ up to a total degree  $d  =300$\footnote{It is in principle possible to compute higher genus zero GW or GV-invariants using\texttt{ CYTools}, but for the present analysis it suffices to work with $d \leq 300.$}. As a preliminary step, we examine the behaviour of the Gromov-Witten (GW) invariants for  across different degrees. As expected \cite{Couso-Santamaria:2016vcc}, there is an exponential growth of the magnitude with the degree $d$. If we perform an exponential fit for the large GW-invariant at each degree for  $d \geq 100 $, we get max($N_d$) $\sim \exp(5.4166 \cdot  d -11.3748)$. This implies that the exponential suppression coming from the instanton action eventually ceases to compensate the growth of the GW-invariants close to the boundary of the large complex structure region. In particular, a typical term in the instanton series at degree $d$ would become $\mathcal{O}(1)$ for Im$(z)\simeq 0.862$.

To monitor the convergence of the instanton expansion, we introduce the following complementary estimators\footnote{ Moreover, one should also compare the absolute difference of the successive instanton corrections between degree $d$ and $d+1$, $|\Delta F_{\rm inst}^{(d+1)} - \Delta F_{\rm inst}^{(d)}|$. }: 
\begin{equation}
    \epsilon_{\rm ratio}^{(d)} = \frac{|\Delta F^{(d+1)}_{\rm inst}|}{|F^{(d)}|}, \quad \epsilon_{\rm relative}^{(d)} = \frac{|\Delta F^{(d+1)}_{\rm inst}|}{|\Delta F^{(d)}_{\rm inst}|}, \,\ \quad  \epsilon_{\rm reference}^{(d)} = \frac{| F^{(100)}_{\rm inst} - F^{(d)}_{\rm inst}|}{|F^{(100)}_{\rm inst}|}.
    \label{eq:control_test}
\end{equation}
Recall that $\Delta F^{(d)}_{\rm inst}$ is the instanton contribution to the prepotential at degree $d$ as opposed to $F^{(d)}_{\rm inst}$, which has all the instanton corrections up to degree $d$. The quantity $\epsilon^{(d)}_{\rm ratio}$ captures the error in the prepotential for neglecting the instanton contributions at degree $d+1$, whereas $\epsilon^{(d)}_{\rm relative}$ compares relative successive instanton contributions between $d$ and $d+1$. Finally, $\epsilon_{\rm reference}^{(d)}$ compares the effect of including all instanton corrections between degree $d$ and $100$. We regard $\epsilon_{\text{reference}}$ as the primary diagnostic, while the remaining estimators provide independent consistency checks.

\begin{figure}[h!]
    \centering
\includegraphics[width=1.0\linewidth]{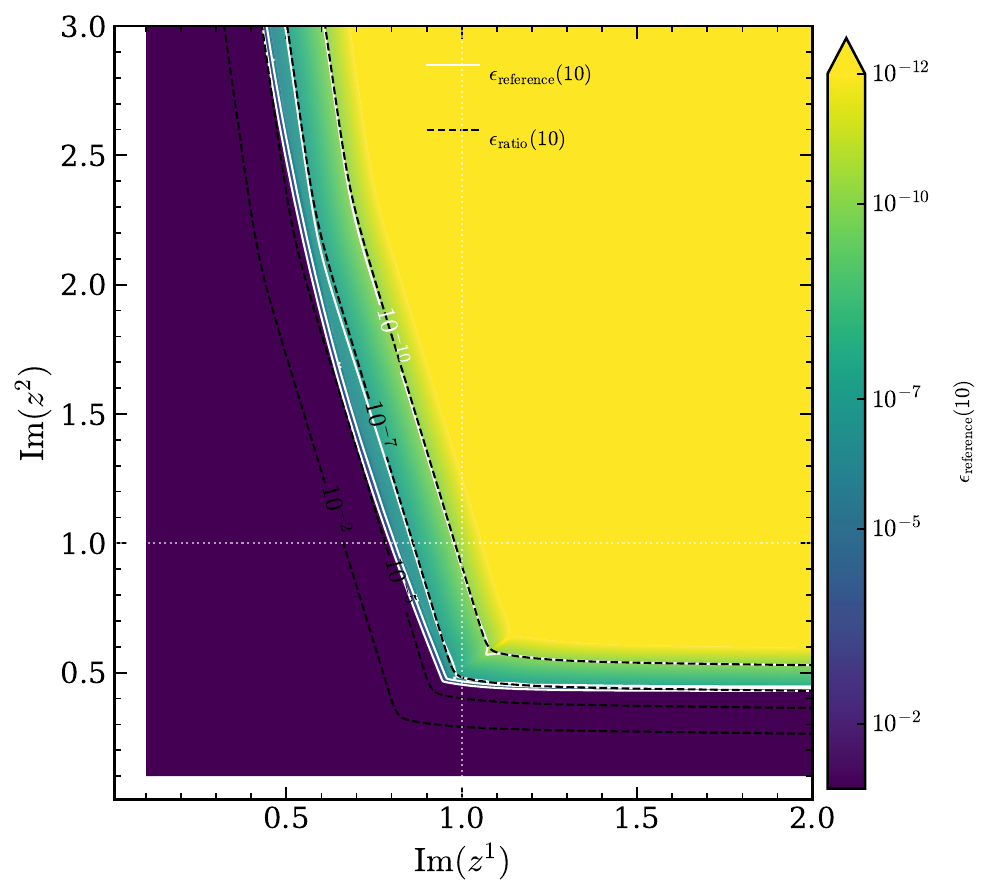}
\caption{Region of convergence in Im$(z^i)$ plane. The dashed black and solid white lines are the contours of $\epsilon_{\rm ratio}$ and $\epsilon_{\rm reference}$, respectively (for the values $10^{-2}, 10^{-5}, 10^{-7}, 10^{-10}$). Vertical and horizontal dashed white lines highlight the Im$(z^i)\geq 1$ region.}
        \label{fig:instanton_converg}
\end{figure}

Fig. \ref{fig:instanton_converg} shows the behaviour of these estimators for the truncation at $d=10$ in the imaginary complex structure moduli plane.\footnote{The real parts of the moduli, after mapping to the fundamental domain with $\mathrm{Re}(z^i)\in[-0.5,0.5]$, contribute only as phases in the instanton expansion and therefore do not affect the region of convergence. Therefore, we omit them.} The dashed black and solid white contours correspond to $\epsilon_{\rm ratio}$ and $\epsilon_{\rm reference}$, respectively, with contour values $10^{-2}$, $10^{-5}$, $10^{-7}$ and $10^{-10}$. The vertical and horizontal dashed white lines indicate $\mathrm{Im}(z^i)=1$.

Three qualitatively different regions can be identified. Deep inside the large complex structure regime (yellow region) both $\epsilon_{\rm ratio}$ and $\epsilon_{\rm reference}$ are extremely small. It corresponds to a region where the truncation is well controlled, and the exponentially suppressed instanton corrections are mostly negligible. 

Closer to the boundary of the LCS patch (green band),  $\epsilon_{\rm reference}$ goes from $10^{-5}$ to $10^{-7}$. Here, the leading instanton correction is large enough to become phenomenologically relevant while the higher contributions remain strongly suppressed, ensuring control over the instanton expansion. This is precisely the region in which the examples of vacua provided in Sec. \ref{sec:basic_examples} are located. The near coincidence of the $\epsilon_{\rm ratio}$ and $\epsilon_{\rm reference}$ contours for the values $10^{-7}$ and $10^{-10}$ indicates that the ratio error provides a reliable measure in this regime. 

Finally, for $ \epsilon_{\rm ratio} \geq 10^{-5}$ (violet region)  the higher-instanton corrections cease to decrease and the series is not convergent.  In this regime, the first instanton correction is no longer representative of the full series since the higher-degree instanton contributions are no longer negligible and the truncated expansion breaks down. The nearly coincident $\epsilon_{\rm reference}$ contours at $10^{-2}$ and $10^{-5}$ illustrate the sharp transition between these two regimes. Another useful diagnostic is the condition $\epsilon_{\rm relative}^{(d)}\sim\mathcal{O}(1)$, which also signals the loss of control over the instanton expansion.

\begin{figure}[h]
    \centering
    \includegraphics[width=1.0\linewidth]{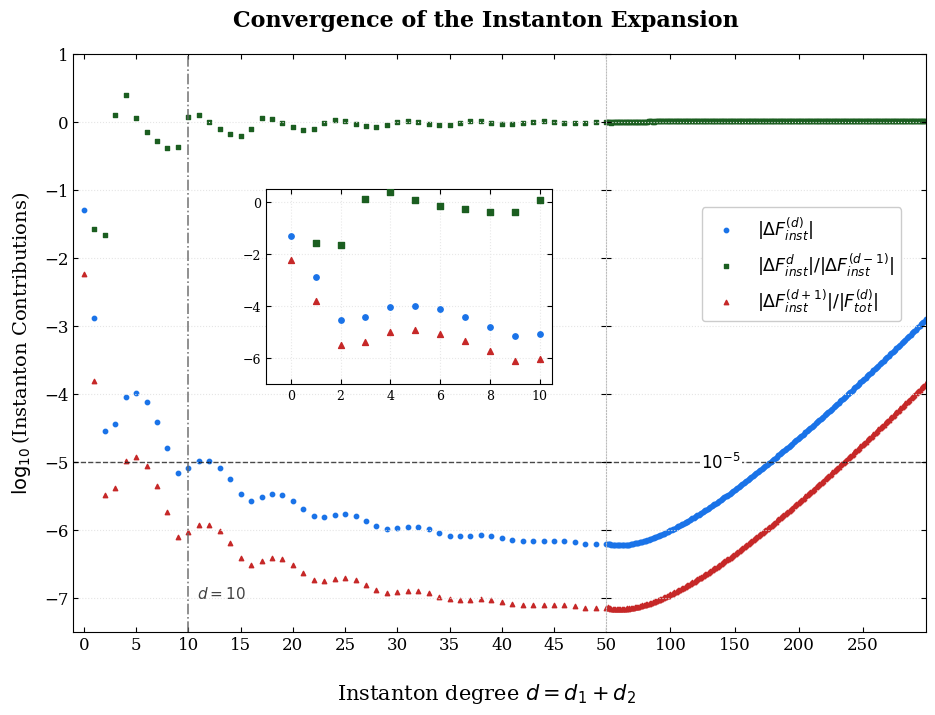}
\caption{Flux vacua behaviour near the boundary of the convergence of the prepotential instanton expansion. The blue, green, and red points correspond to $|\Delta F_{\rm inst}^{(d)}|$, $|\Delta F_{\rm inst}^d|/|\Delta F_{inst}^{(d-1)}|$ and $|\Delta F_{\rm inst}^{(d+1)}|/|F^{(d)}|$. The vertical line at $d=50$ indicates a change of scale along the horizontal axis. 
The successive instanton corrections become less and less important for $d\geq10$ with $\epsilon_{\rm relative}^{(10)} \leq 10^{-6}$; however, as one goes beyond $d\sim50$, these contributions start to be less and less negligible. }
    \label{fig:diverg_eg}
\end{figure}

We now want to highlight the limitations of relying solely on the criterion $\epsilon_{\rm relative}^{(10)}\leq 10^{-5}$ for the instanton series convergence. Let us consider the following flux configuration
\begin{equation}
    f = (-10, -9, -3, 3, -1, -2) \,, \qquad 
    h = (-6, -4, 0, 1, 0, -2) \,,
\end{equation}
with the corresponding perturbative complex structure moduli and axio-dilaton vevs
\begin{align}
    \langle z^1 \rangle &= -0.484804 + 0.866834 \, i \,, \nonumber \\
    \langle z^2 \rangle &= -0.236850 + 0.818167 \, i \,, \\
    \langle \tau \rangle &= \phantom{-}0.485233 + 3.453563 \, i \,. \nonumber
\end{align}
Since both Im$({z^1})$ and Im$({z^2})$ lie close to the boundary of the region of convergence, the vacuum is expected to exhibit interesting behaviour (see Fig.~\ref{fig:diverg_eg}). At the perturbative level, the vacuum solves the F-term equations and the convergence condition $\epsilon_{\rm ratio}^{(10)} \leq 10^{-6}$. One might therefore conclude that the truncation at degree $d=10$ provides an accurate approximation of the non-perturbative solution. However, incorporating higher-degree instanton corrections, the expansion ceases to be under control at around $d = 50$, where all higher instantons start becoming non-negligible. Consequently, the perturbative vacuum must lie outside the region of convergence for $F_{\rm inst}$ where the instanton expansion is not under control. 

\begin{figure}[h!]
    \centering
    \includegraphics[width=1.0\linewidth]{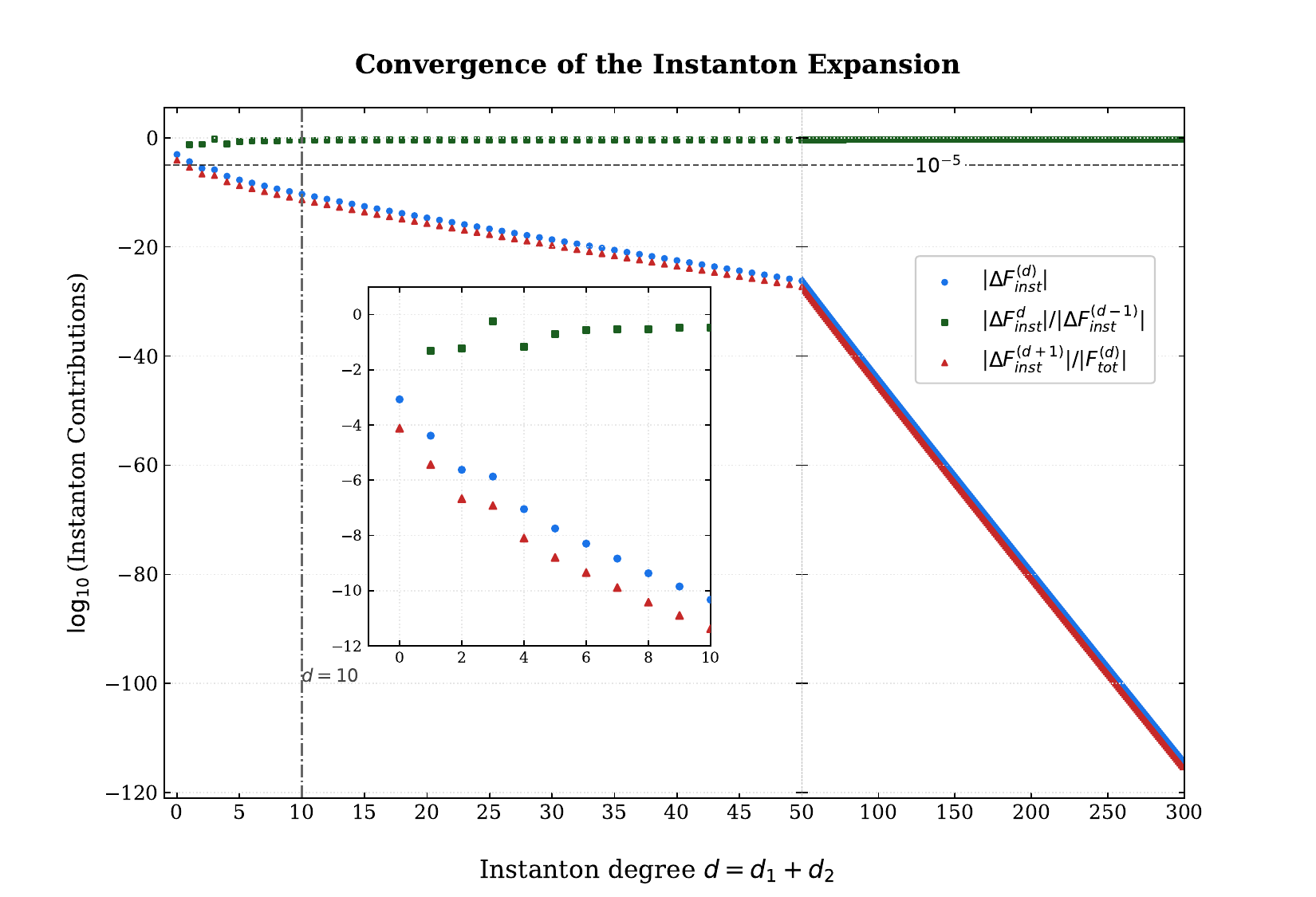}
\caption{Flux vacua behaviour near the boundary of the convergence of the prepotential instanton expansion. The blue, green, and red points correspond to $|\Delta F_{\rm inst}^{(d)}|$, $|\Delta F_{\rm inst}^d|/|\Delta F_{\rm inst}^{(d-1)}|$ and $|\Delta F_{\rm inst}^{(d+1)}|/|F^{(d)}|$. The vertical line at $d=50$ indicates a change of scale along the horizontal axis. 
The successive instanton corrections decay for higher and higher degree and noticeably $\epsilon_{\rm relative}^{(10)} \leq 10^{-12}$.  }
\label{fig:converg_eg}
\end{figure}

This perturbative flux vacuum, upon including $F_{\rm inst}$ up to degree $10$, shifts further in the large complex structure, albeit leaving $U$. After the monodromy map, $\text{Sp}(6,\mathbb{Z})$, and $\text{SL}(2,\mathbb{Z})$, is physically equivalent to the flux:
\begin{equation}
    f = (9, 2, -2, 0, -1, 4) \,, \qquad 
    h = (-3, -1, 0, 1, 0, -1) \,.
\end{equation}
with the moduli:
\begin{align}
    \langle z^1 \rangle &= -0.142078 + 1.250608 \, i \,, \nonumber \\
    \langle z^2 \rangle &= -0.431737 + 0.608257 \, i \,, \\
    \langle \tau \rangle &= -0.220496 + 1.991678 \, i \,. \nonumber
\end{align}

After incorporating instanton corrections, the minimum shifts into a region where instanton corrections are under control, i.e. become rapidly unimportant. Fig.~\ref{fig:converg_eg} contains the corresponding behaviour of various tests as discussed in \eqref{eq:control_test}. As can be seen, the successive instanons are highly suppressed. 
 
\bibliographystyle{utphys}
\bibliography{biblio}

\end{document}